\documentclass[aps,prl,amsmath,amssymb,reprint,superscriptaddress,nofootinbib]{revtex4-2}

\usepackage{graphicx}
\usepackage{dcolumn}
\AtBeginDocument{\usepackage{booktabs}}
\usepackage{units}
\usepackage{hyperref}

\usepackage{color}
\definecolor{red}{RGB}{255,0,0}

\usepackage{comment}
\usepackage{amsmath,amsthm,amssymb}
\usepackage{physics}
\usepackage{dsfont} 

\usepackage{xcolor}

\providecommand{\rang}{\rangle}
\providecommand{\lang}{\langle}

\newtheorem{lemma}{Lemma}
\newtheorem*{lemma*}{Lemma}

\newcommand{\cN}{\mathcal{N}}
\newcommand{\cO}{\mathcal{O}}

\newcommand{\cU}{\mathcal{U}}

\usepackage[normalem]{ulem}
\providecommand{\ignore}[1]{}
\providecommand{\auedit}[1]{#1}
\newif\ifcmnt
\cmnttrue
\ifdefined\cmntsoff\cmntfalse\fi
\ifcmnt
    \providecommand{\aucmnt}[1]{#1}
    \providecommand{\mkcolor}[1]{\color{brown}{#1}}

\else
    \providecommand{\aucmnt}[1]{}
    \providecommand{\mkcolor}[1]{{#1}}

\fi

\newcommand{\MKc}[1]{\aucmnt{{\mkcolor{[MK:} \color{gray} #1\mkcolor{]}}}}
\newcommand{\MKs}[1]{\auedit{{\mkcolor{\sout{#1}}}}}

\usepackage{enumitem}

\begin{document}
	
\preprint{}

\newcommand{\thetitle}{Coherent coupling and non-destructive measurement of trapped-ion mechanical oscillators}
\title{\thetitle}

\author{Pan-Yu Hou}
\email[]{panyu.hou@colorado.edu}
\affiliation{National Institute of Standards and Technology, 325 Broadway, Boulder, CO 80305, USA}
\affiliation{Department of Physics, University of Colorado, Boulder, CO 80309, USA}

\author{Jenny J.~Wu}
\affiliation{National Institute of Standards and Technology, 325 Broadway, Boulder, CO 80305, USA}
\affiliation{Department of Physics, University of Colorado, Boulder, CO 80309, USA}

\author{Stephen D.~Erickson}
\affiliation{National Institute of Standards and Technology, 325 Broadway, Boulder, CO 80305, USA}
\affiliation{Department of Physics, University of Colorado, Boulder, CO 80309, USA}
\altaffiliation[]{Current address: Quantinuum, 303 South Technology Court, Broomfield, Colorado 80021, USA}

\author{Daniel C.~Cole}
\affiliation{National Institute of Standards and Technology, 325 Broadway, Boulder, CO 80305, USA}
\altaffiliation[]{Current address: ColdQuanta, Inc., Boulder, Colorado 80301, USA}

\author{Giorgio Zarantonello}
\affiliation{National Institute of Standards and Technology, 325 Broadway, Boulder, CO 80305, USA}
\affiliation{Department of Physics, University of Colorado, Boulder, CO 80309, USA}
\author{Adam D.~Brandt}
\affiliation{National Institute of Standards and Technology, 325 Broadway, Boulder, CO 80305, USA}

\author{Shawn Geller}
\affiliation{National Institute of Standards and Technology, 325 Broadway, Boulder, CO 80305, USA}
\affiliation{Department of Physics, University of Colorado, Boulder, CO 80309, USA}

\author{Alex Kwiatkowski}
\affiliation{National Institute of Standards and Technology, 325 Broadway, Boulder, CO 80305, USA}
\affiliation{Department of Physics, University of Colorado, Boulder, CO 80309, USA}

\author{Scott Glancy}
\affiliation{National Institute of Standards and Technology, 325 Broadway, Boulder, CO 80305, USA}

\author{Emanuel Knill}
\affiliation{National Institute of Standards and Technology, 325 Broadway, Boulder, CO 80305, USA}

\author{Andrew C.~Wilson}
\affiliation{National Institute of Standards and Technology, 325 Broadway, Boulder, CO 80305, USA}
\author{Daniel H.~Slichter}
\affiliation{National Institute of Standards and Technology, 325 Broadway, Boulder, CO 80305, USA}
\author{Dietrich Leibfried}
\email[]{dietrich.leibfried@nist.gov}
\affiliation{National Institute of Standards and Technology, 325 Broadway, Boulder, CO 80305, USA}

\date{\today} 
\begin{abstract}  
\noindent 
%
Precise quantum control and measurement of multiple harmonic oscillators, such as modes of the electromagnetic field in a cavity or of mechanical motion, are key for their use as quantum platforms.
The motional modes of trapped ions can be individually controlled and possess good coherence properties, but to date have lacked high-fidelity two-mode operations and non-destructive motional state measurement.
Here, we demonstrate the coherent exchange of single motional quanta between spectrally separated harmonic motional modes of a trapped ion crystal. The timing, strength, and phase of the coupling are controlled through an oscillating electric potential with suitable spatial variation. 
Coupling rates that are much larger than decoherence rates enable demonstrations of high-fidelity quantum state transfer and beam-splitter operations, entanglement of motional modes, and Hong–Ou–Mandel-type interference. Additionally, we use the motional coupling to enable repeated non-destructive projective measurement of a trapped ion motional state.
Our work enhances the suitability of trapped-ion motion for continuous-variable quantum computing and error correction, and may provide opportunities to improve the performance of motional cooling and motion-mediated entangling interactions. 
%
\end{abstract}
\maketitle
Harmonic oscillators (HOs) are ubiquitous in models of nature, and many elementary phenomena are described by the interaction between HOs. The high-dimensional Hilbert space of HOs can be used to encode and process quantum information, given control operations of sufficient fidelity~\cite{feynman2018simulating, leibfried2002trapped, porras2004bose, bermudez2011synthetic,furusawa1998unconditional,braunstein1998quantum,ralph1999continous,van2000multipartite,chuang1997bosonic,knill2001scheme,gottesman2001encoding,braunstein2005quantum, kok2007linear}. Quantum error-correcting codes can take advantage of this large Hilbert space to reduce hardware requirements relative to codes based on two-level systems~\cite{lloyd1999quantum,chuang1997bosonic,gottesman2001encoding, braunstein1998error, michael2016new}.
Individual control and coherent coupling of HOs near or in the quantum regime have been demonstrated across many physical systems~\cite{brown2011coupled, harlander2011trapped, wilson2014tunable,toyoda2015hong,gorman2014two,hong1987measurement,rauschenbeutel2001controlled,groblacher2009observation,teufel2011circuit,wang2011deterministic,zakka2011quantum,verhagen2012quantum,shalabney2015coherent,gao2018programmable,capmany2020programmable,palomaki2013coherent,kotler2021direct, chapman2022high}.
Among them, modes of trapped-ion motion~\cite{wineland1998experimental,leibfried2003quantum} and of superconducting cavities in the circuit quantum electrodynamics (cQED) architecture~\cite{heeres2017implementing,gao2018programmable,gao2019entanglement} exhibit long coherence times and can be individually controlled with high fidelity, as required for continuous-variable quantum information processing (QIP). Bosonic QEC beyond the break-even point has recently been demonstrated in cQED systems~\cite{sivak2023real,ni2023beating}.

Continuous-variable QIP in trapped-ion systems is currently limited by the lack of controllable direct coupling between motional modes for two-mode operations, and of methods for non-destructive measurement of motional states. 
The latter is a prerequisite for QEC based on error syndrome measurement and has been a long-standing shortcoming~\cite{wineland1998experimental,fluhmann2019encoding,de2022error} because photon recoil during atomic state fluorescence readout stochastically alters the motion of the addressed ion and scrambles any information stored in its motional state.
Previous work has relied on probabilistic state preparation and measurement via post-selection of ``dark" readout events (without photon scattering)~\cite{kienzler2016obvservation,fluhmann2019encoding,gan2020hybrid}, or on dissipative state preparation that does not involve detection~\cite{de2022error}.

Motional modes of ions have been coupled through the ions' internal (electronic) states using lasers, but this does not work for ions that are not accessible to lasers or do not participate in modes to be coupled~\cite{jost2009entangled,wolf2016non,gan2020hybrid}. Motional modes of ions in separate potential wells have also been resonantly coupled by the Coulomb interaction, with the coupling strength limited by the ion-ion spacing and restricted control~\cite{brown2011coupled,harlander2011trapped,wilson2014tunable,toyoda2015hong}, and modulation of the trap potential has been used to couple two motional modes of an electron cloud or a single trapped ion~\cite{wineland1975principles,gorman2014two}.
%

Here, we demonstrate direct coupling\textemdash with controllable timing, strength, and phase\textemdash between two motional modes in a linear, mixed-species ion crystal. The duration of state exchange can be much shorter than motional coherence times, enabling high-fidelity two-mode operations. The coupling is generated by a suitable spatially varying oscillating electric potential, can be used for crystals of arbitrary size, and is independent of ion internal structure, making it applicable to any Coulomb crystal, including those containing molecular ions, highly charged ions, or other species that lack easily accessible transitions between internal states. More generally, well-controlled mode coupling can be used to improve cooling and quantum logic operations for ion-based QIP, timekeeping, and quantum sensing applications~\cite{Modecouplingforcooling}. 

\begin{figure*}
    \centerline{\includegraphics[width=1\textwidth]{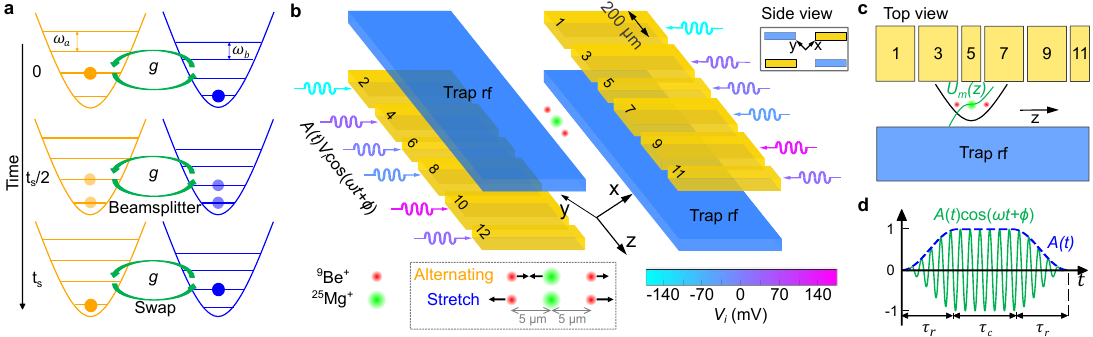}}\label{fig:fig1}
    \caption{\textbf{Coupled quantum mechanical oscillators.}  \textbf{a}, Illustration of quantum state transfer between two coupled HOs. Coherent coupling (green) fully exchanges states up to a phase at $t_{s}$ and creates two-mode entanglement at $t_{s}$/2, constituting a beamsplitter operation (see Eq.~(\ref{Eq:CoupH}) and Methods).
    \textbf{b,c}, Three-dimensional perspective, side view (inset; \textbf{b}) and top view (\textbf{c}) of trapping zone in segmented Paul trap (not to scale) with radio-frequency voltages applied to two electrodes (blue, ‘trap rf’).  . A $^{9}$Be$^{+}$-$^{25}$Mg$^{+}$-$^{9}$Be$^{+}$ crystal is confined along axial direction $z$ in a harmonic potential (solid black line in top view). Two of three axial normal modes, ``Alternating'' at 3.66\,MHz and ``Stretch'' at 3.38\,MHz (mode participation vectors $\xi$ for all ions visualized as arrows in the dashed box at the bottom), are coupled by an oscillating electric potential $U_\mathrm{mod}({z,t}) =  U_m(z)A(t)\cos(\omega t+\phi)$. The coordinate origin is at the Mg$^+$ ion, and the spatial dependence of $U_m(z) \propto z^{3}$ is visualized by the green curve in \textbf{c}. The potential is generated from synchronized oscillating drives $V_{i} A(t) \cos(\omega t+\phi)$ for $i\in\{1,...,12\}$ applied to twelve control electrodes (gold). The $V_i$ values represented by the color of the wavy arrows correspond to a coupling rate of $2g_{0}=2\pi\times 5.1$~kHz when $A(t)=1$.
    \textbf{d}, Coupling pulse shape in experiments. The green oscillating line represents temporal dependence of $U_\mathrm{mod}({z,t})$ with amplitude envelope $A(t)$ (blue dashed line). For a coupling pulse with non-zero duration, $A(t)$ ramps up from zero to one in $\tau_{r}$\,=\,20\,$\mu$s, stays constant for $\tau_{c}$ and ramps back to zero in $\tau_{r}$. The pulse area is equal to that of a square pulse of amplitude one and duration $\tau=\tau_{r}+\tau_{c}$.
    }
\end{figure*}

We also use mode coupling to perform repeated non-destructive measurements of ion motion. For ion crystals with a mirror symmetry around the center ion, this ion must have zero amplitude in all modes with an even motion pattern. Even modes are therefore ``protected'' from recoil of the center ion. By the same argument, the modes that do not suffer recoils from the center ion are inaccessible through that ion. Mode coupling enables us to transfer information about a motional state onto the center ion, swap the state of motion into a protected mode, and then read out the center ion (thus learning information about the motional state of interest) without destroying the motional state due to photon recoils. This protocol can then be repeated to achieve greater confidence in the result, or to realize repeated rounds of syndrome measurement for a bosonic error correcting code.
%


A linear string of $N$ ions confined in a three-dimensional harmonic potential exhibits $3 N$ normal modes of collective motion that can be treated as uncoupled HOs~\cite{wineland1998experimental,james1998quantum}.
Consider two normal modes $a$ and $b$ at frequencies $\omega_{a}$ and $\omega_{b}$, with ladder operators $\hat{a}$ and $\hat{b}$. 
Their coupling can be described by Hamiltonian
\begin{equation}\label{Eq:CoupH}
H= \hbar g \left(e^{i\phi}\hat{a} \hat{b}^{\dagger} + e^{-i\phi}\hat{b} \hat{a}^{\dagger}\right), 
\end{equation}
where $2\pi \hbar$ is Planck's constant, $\hbar g$ is the coupling energy, and $\phi$ is the coupling phase. This coupling leads to state exchange between modes $a$ and $b$, as illustrated in Fig.~\ref{fig:fig1}\textbf{a}. 
Ideally, the timing, strength $g$, and phase $\phi$ of the coupling can be well controlled. 

To couple modes, we add to the existing confining trap potential an oscillating and spatially-varying electric potential modulation of the form\begin{equation}\label{Eq:CoupPot}
U_\mathrm{mod}(\boldsymbol{r},t) =U_m(\boldsymbol{r})A(t)\cos(\omega t+\phi), 
\end{equation}
with $\omega \approx |\omega_{a}-\omega_{b}|$ and $0 \leq A(t) \leq 1$. The smooth envelope $A(t)$ (blue dashed line in Fig.~\ref{fig:fig1}\textbf{d}) evolves slowly compared to $2\pi/\omega$ to avoid sudden perturbations of the trapping potential. Modes are coupled by curvature terms
\begin{equation}
\alpha_{n} = \frac{\partial ^{2} U_m}{\partial i_{a}\partial i_{b}}\bigg|_{\boldsymbol{r}=\boldsymbol{r_{n,0}}}
\end{equation}
in the expansion of $U_m(\boldsymbol{r})$ around the $n$th ion's equilibrium position $\boldsymbol{r}_{n,0}$ along the mode directions $i_a, i_b \in \{x,y,z\}$. The coupling strength $g_0$ is a sum over contributions from each ion
\begin{equation}\label{Eq:CoupRat}
g_0 = \sum_{n=1}^{N}g_{n} =\sum_{n=1}^{N} \left(\frac{Q_{n}}{4M_{n}\sqrt{\omega_{a}\omega_{b}}}\times\alpha_{n}\xi^{(i_a)}_{n,a}\xi^{(i_b)}_{n,b}\right), 
\end{equation} 
where $Q_{n}$, $M_{n}$,  $\xi^{(i_a)}_{n,a}$ and $\xi^{(i_b)}_{n,b}$ denote the charge, mass and participation in modes $a$ and $b$ of the $n$th ion. The participation is defined as the $n$th ion's component of the normalized eigenvector of a given normal mode. After transforming into the interaction picture and neglecting fast-rotating terms (see Methods), the Hamiltonian associated with the modulation in Eq.~(\ref{Eq:CoupPot}) becomes Eq.~(\ref{Eq:CoupH}) with $g(t)=A(t) g_0$. A suitable choice of $U_m(\boldsymbol{r})$ sets the signs of the products $\alpha_n \xi^{(i_a)}_{n,a}\xi^{(i_b)}_{n,b}$ so that the  $g_{n}$ add constructively.

We trap $^9$Be$^+$ and $^{25}$Mg$^+$ ions in a segmented linear Paul trap shown in  Fig.~\ref{fig:fig1}\textbf{b} and \ref{fig:fig1}\textbf{c} (see also~\cite{blakestad2010transport}). We denote the linear trap axis as $z$ and the trap potential ellipsoid radial principal axes as $x$ and $y$. To produce the coupling potential $U_\mathrm{mod}(\boldsymbol{r},t)$, we apply voltages of the form $V_{i} A(t) \cos(\omega t+\phi)$, $i\in\{1,...,12\}$, to the twelve electrodes closest to the ions, using simulations of the potentials created by each electrode at the ion positions to determine the desired $V_{i}$. In each experiment, $^{9}$Be$^+$ is prepared in $\left|\downarrow\right\rangle_{B}\equiv \,^{2}S_{1/2} \left|F=2,m_F=2\right\rangle_B$ and $^{25}$Mg$^+$ in $\left|\downarrow\right\rangle_M \equiv \,^{2}S_{1/2} \left|F=3,m_F=3\right\rangle_M$ by optical pumping; transitions to the other qubit states $\left|\uparrow\right\rangle_B \equiv \,^{2}S_{1/2} \left|F=1,m_F=1\right\rangle_B$ and $\left|\uparrow\right\rangle_M\equiv \,^{2}S_{1/2}\left|F=2,m_F=2\right\rangle_M$ are driven by microwave magnetic fields or Raman laser beams. Motional mode information is mapped into internal states using sideband transitions~\cite{wineland1998experimental} and read out by state dependent fluorescence. Ions in the ``bright'' states $\left|\downarrow\right\rangle_B$ and $\left|\downarrow\right\rangle_M$ scatter thousands of photons during readout, of which approximately 30 photons on average are detected, while all other hyperfine states (``dark'' states) scatter zero or a few photons. Further details are provided in the Supplementary Material.
 
We demonstrate the essential features of the coupling on the ``Alternating'' ($\sim$3.66\,MHz, subscript $A$) and ``Stretch" ($\sim$3.38\,MHz, subscript $S$)  axial modes of a $^{9}$Be$^{+}$-$^{25}$Mg$^{+}$-$^{9}$Be$^{+}$ mixed-species crystal. The participations of ions in each mode are represented by black arrows in the lowest panel of Fig.~\ref{fig:fig1}\textbf{b}. The Mg$^{+}$ ion does not contribute to $g_0$ because it has no participation in the Stretch mode. A cubic oscillating potential $U_\mathrm{mod}({z,t}) = A(t) U_m(z) \cos(\omega t+\phi)\propto z^{3}$ yields opposite $\alpha_n$ for the two Be$^{+}$ ions, so that the $g_n$ add constructively.

\begin{figure*}
    \centerline{\includegraphics[width=1\textwidth]{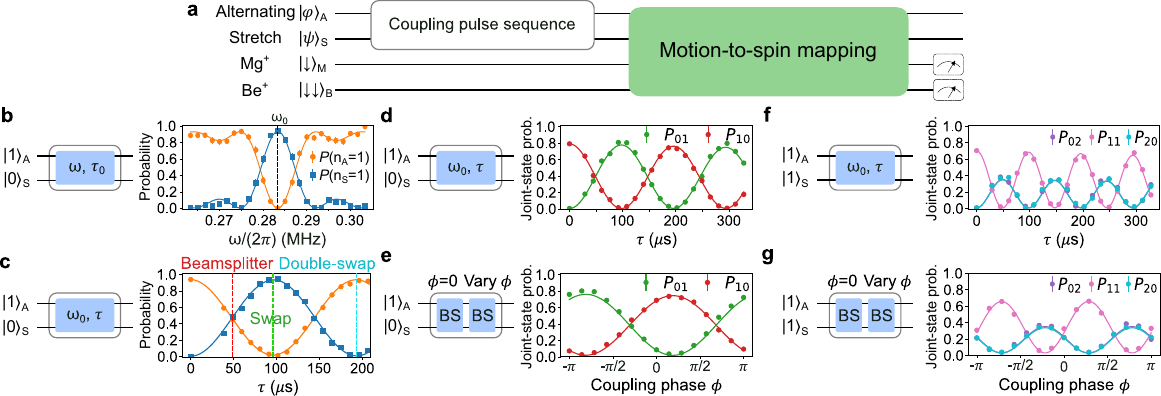}}
    \caption{\textbf{Coherent coupling dynamics.}  \textbf{a}, Experimental sequence to characterize Alternating-Stretch coupling. Modes are prepared by Raman laser interactions, then coupled by the pulse sequence in the white box. Motional state probabilities are then mapped onto internal states of one or both ion species by further Raman laser interactions (green box, see details in text) followed by state-dependent fluorescence detection. Initial motional states and pulse sequences (one blue box per coupling pulse) are indicated in \textbf{b}-\textbf{g}.
    Lines in \textbf{b}, \textbf{c} are fits to data, while lines in \textbf{d}-\textbf{g} are from numerical simulation using experimental parameters. 
    \textbf{b}, With initial state $|1\rangle_A |0\rangle_S$, $P(n_A=1)$ is high unless the coupling frequency $\omega$ is tuned near resonance, where instead the probability of the phonon being found in the Stretch mode is high (blue squares). 
    \textbf{c}, With the coupling on resonance, a single phonon is coherently swapped between the two modes as coupling time $\tau$ increases. Vertical dashed lines indicate pulse durations for beamsplitter (BS), swap, and double-swap operations. 
    %
    %
    \textbf{d-g}, Probability of finding certain states as exchange duration $\tau$ is varied for initial states $\ket{1}_{A}\ket{0}_{S} $(\textbf{d}) and $\ket{1}_{A}\ket{1}_{S} $ (\textbf{f}), or as coupling phase is varied for initial states $\left|1\right\rangle_{A} \left|0\right\rangle_{S}$ (\textbf{e}) and $\left|1\right\rangle_{A} \left|1\right\rangle_{S}$ (\textbf{g}) . Labels $P_{a s}$ indicate there are $a$ phonons in the Alternating mode, and $s$ phonons in the Stretch mode).
    Results in (\textbf{d, e}) and (\textbf{f, g}) verify two-mode entanglement generated by a BS.
    Results in (\textbf{f, g}) correspond to Hong-Ou-Mandel-type interference between two phonons at different frequencies. Each data point was obtained from 300 experiments in \textbf{b,c} and from 1,000 experiments in \textbf{d-g}, with a 68\% confidence error bar. prob., probability. 
    }
    \label{fig:fig2}
\end{figure*}

We calibrate the optimal modulation frequency $\omega_{0}$ and the coupling strength $g_0$ by preparing the two modes in the state $\left|1\right\rangle_{A}\left|0\right\rangle_{S}$. A coupling pulse alters the probability $P(n_{A/S}=1)$ of which mode contains the motional quantum (phonon). When scanning the modulation frequency $\omega$ (Fig.~\ref{fig:fig2}\textbf{b}) around $\omega_A-\omega_S$ with fixed coupling pulse duration $\tau_{0}\approx$\,100$\,\mu$s, the probability $P$($n_{A}=1)$ is reduced to near zero coincident with an increase of the probability $P$($n_{S}=1)$. With the drive frequency fixed at $\omega_0$ we scan the pulse duration $\tau$, observing $P$($n_{A}=1)$ and $P$($n_{S}=1)$ oscillating out of phase at frequency $\Omega_{c}=2g_{0}\approx 2\pi\times5.1$~kHz, as shown in Fig.~\ref{fig:fig2}\textbf{c}. The single phonon is swapped into the Stretch mode at $t_{s}\approx$\,100$\,\mu$s and transferred back to the Alternating mode (``double-swap'') at $2t_{s}$. 
The loss of population from the $n=\{0,1\}$ subspace per swap is estimated to be about 0.5\%, mainly due to motional heating into states with $n>1$ (see Supplementary Material).
We can attain a maximum coupling frequency $\Omega_{c}$ of about $2\pi\times18$~kHz, limited by the drive electronics (see Methods). 

The dynamics of coherent coupling are characterized by measuring correlations between coupled motional modes. For each of the two modes, we map the amplitudes of specific number states onto the internal states of one ion species, with the other mode mapped to the other ion species. We can then perform joint measurement of the mapped motional mode information for both modes in a single experimental trial, using species-resolved state-dependent fluorescence readout (see Supplementary Material).
The interaction given by Eq.~(\ref{Eq:CoupH}) ideally conserves the total number $\mathcal{N}$ of phonons in both modes. Experimentally, only populations of states with the same $\mathcal{N}$ were found to be substantial, shown as dots in Fig.~\ref{fig:fig2}\textbf{d}-\textbf{g} along with simulations (lines)~\cite{Johansson2013} based on experimental parameters. The state preparation and measurement errors are larger in Fig.~\ref{fig:fig2}\textbf{d}-\textbf{g} compared to those in Fig.~\ref{fig:fig2}\textbf{b,c} because more complex pulse sequences are used during state preparation and readout (see Supplementary Material).

%
In Fig.~\ref{fig:fig2}\textbf{d}, with initial state $\left|1\right\rangle_{A} \left|0\right\rangle_{S}$ ($\mathcal{N}=1$) the population swaps into $\left|0\right\rangle_{A} \left|1\right\rangle_{S}$ and we observe two anti-correlated sinusoidal population oscillations with similar amplitude.
At $t_{BS}$\,$\approx$\,50\,$\mu$s, the coupling pulse realizes a beamsplitter (BS) operation $U_{BS}= \exp[i(\pi/4)(\hat{a} \hat{b}^{\dagger} + \hat{b} \hat{a}^{\dagger})]$, which we expect to generate an entangled state $(\left|1\right\rangle_{A}\left|0\right\rangle_{S}+\left|0\right\rangle_{A}\left|1\right\rangle_{S})/\sqrt{2}$.
We observe approximately equal population in $\left|0\right\rangle_{A}\left|1\right\rangle_{S}$ and $\left|1\right\rangle_{A}\left|0\right\rangle_{S}$ at $t_{BS}$ and verify the coherence between these two components by
performing a phonon interferometry experiment consisting of two BS operations with variable phase difference $\phi$ (Fig.~\ref{fig:fig2}\textbf{e}), showing the coherence of the entangled state generated by the first BS pulse. We estimate an 84\% confidence lower bound of the average fidelity in the one-phonon subspace of the beamsplitter operation to be 97.9\% (see Methods).
The initial state $\left|1\right\rangle_{A} \left|1\right\rangle_{S}$ ($\mathcal{N}$=2) evolves into  $\left|0\right\rangle_{A} \left|2\right\rangle_{S}$ and $\left|2\right\rangle_{A} \left|0\right\rangle_{S}$ with nearly equal population at $t_{BS}$, while the population in $\left|1\right\rangle_{A} \left|1\right\rangle_{S}$ is reduced almost to zero by destructive interference (Fig.~\ref{fig:fig2}\textbf{f}). This behavior is analogous to Hong–Ou–Mandel interference~\cite{hong1987measurement} but here phonons at different frequencies interfere. An entangled  state of the form $(\left|2\right\rangle_{A}\left|0\right\rangle_{S}+\left|0\right\rangle_{A}\left|2\right\rangle_{S})/\sqrt{2}$ is generated at $t_{BS}$, and phase coherence of this state is also verified (Fig.~\ref{fig:fig2}\textbf{g}). 
Additional conditional probabilities were tracked and are shown in Figs.~\ref{fig:fig7}-\ref{fig:fig10} in Supplementary Material.
%
\begin{figure*}
    \centerline{\includegraphics[width=1\textwidth]{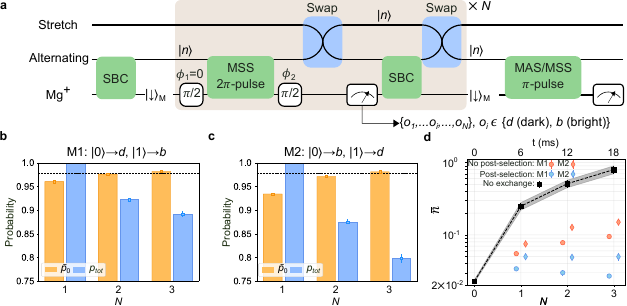}}
    \caption{\textbf{Repeated interrogation of a near ground state thermal distribution of trapped-ion motion.} \textbf{a}, Pulse sequence for non-destructively distinguishing number states $\{\left|0\right\rangle$,\,$\left|1\right\rangle\}$. SBC: sideband cooling, MAS/MSS: motion adding/subtracting sideband. See description in the main text. 
    \textbf{b} and \textbf{c}, Repeated measurement outcomes. With increasing $N$, the post-selected probability $\Tilde{p}_{0}$ (orange bars) of determining the motional state to be $\left| 0\right\rangle$ approaches the independently calibrated population $p_{0}$ (dashed line) in the initial motional state distribution. The similarity between mapping M1 (\textbf{b}), where few fluorescence photons are scattered detecting $\left| 0\right\rangle$, and mapping M2 (\textbf{c}), where thousands of fluorescence photons are scattered detecting $\left| 0\right\rangle$, shows the robustness of the protected state to photon recoil. The blue bars show $p_{tot}$, the total probability of post-selection (see main text).
    \textbf{d}, Alternating mode mean occupation number ($\bar{n}$) post-selected on all $N$ outcomes being $|0\rangle$ (blue symbols) is lower than corresponding $\bar{n}$ with no post-selection (red symbols). ``No exchange'' $\bar{n}$ (black squares) are measured after applying a delay with the duration of $N$ measurement blocks without swapping into the Stretch mode. 
    Each data point with a 68\% confidence error bar was obtained from 6,000 experiments except for M2 $N$=3, which was obtained from 2,000 experiments. Data points and bars are laterally offset from $N$ values for legibility and error bars for some points are smaller than plot symbols.
    }
    \label{fig:fig3}
\end{figure*}

Atomic motion 
can be significantly perturbed by photon recoil during fluorescence-based internal state readout. By limiting photon recoil to a certain ion $j$ that does not participate in a mode $a$, \textit{i.e.} $\xi^{(i_a)}_{j,a}=0$, this mode is protected and unperturbed by photon scattering from ion $j$. 
However, if ion $j$ does not participate in mode $a$, the mode $a$ can also not be directly coupled with internal states of ion $j$. 
Mode coupling enables the state of mode $a$ to be swapped into a suitable mode $b$ which can be coupled to the internal states of ion $j$,  $\xi^{(i_b)}_{j,b}\neq 0$.
Subsequently, information about the state of mode $b$ can be coherently mapped to the internal states of ion $j$ in such a way that the motional state is not altered. 
After swapping the motional state back into mode $a$, readout of ion $j$ yields the encoded state information, while preserving the state of mode $a$ to a high degree aside from measurement projection.

As proof of principle, we implement a protocol to non-destructively distinguish number states $\left|n\right\rangle \in\{\left|0\right\rangle$,\,$\left|1\right\rangle\}$ of ion motion using the circuit shown in Fig.~\ref{fig:fig3}\textbf{a}. 
While $\left|n\right\rangle$ is in the Alternating mode, information can be mapped onto the Mg$^{+}$ internal states with a Cirac-Zoller-type sequence \cite{ciraczoller95} (schematically shown in grey box, see Supplementary Material for details) consisting of a motion-subtracting-sideband (MSS) 2$\pi$ pulse surrounded by two carrier $\pi/2$ pulses. The MSS pulse ideally has no effect on $\left|0\right\rangle_A\left|\downarrow\right\rangle_M$ but transforms $\left|1\right\rangle_A\left|\downarrow\right\rangle_M$ into $-\left|1\right\rangle_A\left|\downarrow\right\rangle_M$. The two $\pi/2$ pulses turn this $\left | n\right\rangle_A$-dependent phase shift into an $\left | n\right\rangle_A$-dependent population difference in the Mg$^{+}$ internal states $\left|\uparrow\right\rangle_M$ and $\left|\downarrow\right\rangle_M$. 
The state $\left|n\right\rangle_A$ is then swapped into $\left|n\right\rangle_S$ of the Stretch mode that does not couple to the Mg$^{+}$, followed by Mg$^{+}$ fluorescence detection, during which the state $\left|\downarrow\right\rangle_M$ scatters thousands of photons (outcome bright ``$b$''), while the state $\left|\uparrow\right\rangle_M$ scatters zero or very few photons (outcome dark ``$d$''). We can repeat the mapping and measurement sequence shown in the grey box by sideband cooling the In-Phase and Alternating modes near the ground state and swapping $\left|n\right\rangle_S$ back into the Alternating mode. To test the resilience of the motional state to photon recoil during fluorescence detection, we perform repeated state measurements with two opposite state mappings M1 and M2, where M1:~$\{\left|0\right\rangle_A \rightarrow d , \, \left|1\right\rangle_A \rightarrow b\}$ and M2:~$\{\left|0\right\rangle_A \rightarrow b , \, \left|1\right\rangle_A \rightarrow d\}$. 

In the experiments, the Alternating mode is cooled to a thermal distribution with an average occupation $\bar{n}$\,=\,0.023(1), with probability $p_0$=0.978(1) of being in $\left|0\right\rangle_A$, $p_1$=0.022(1) in $\left|1\right\rangle_A$, and almost zero in higher number states. We repeat the motional state mapping and Mg$^+$ readout up to three times and obtain a series of outcomes $\{o_{1},...,o_{i},...,o_{N}\}$ with $o_{i} \in \{d, b\}$, $i$=1,...,\,$N$. We post-select only those trials where all $N$ outcomes are the same (all $d$ or all $b$) to improve state discrimination. The post-selected relative frequencies of declaring $\left|0\right\rangle$ and $\left|1\right\rangle$ are defined as $\Tilde{p}_{0}= p(\{d\}_{N})/p_{tot}$ for M1 or $ \Tilde{p}_{0}=p(\{b\}_{N})/p_{tot}$ for M2, with $\Tilde{p}_{1}=1-\Tilde{p}_{0}$. Here $p(\{d\}_{N})$ and $p(\{b\}_{N})$ are the probabilities of all $N$ outcomes being $d$ or $b$ respectively, and $p_{tot}=p(\{d\}_{N})+p(\{b\}_{N})$ is the total probability of post-selection. 

The state after $N$ rounds of interrogation can be characterized independently by applying a $\pi$ pulse on the motion-adding-sideband (MAS) or MSS transition of the Alternating mode, followed by another Mg$^{+}$ fluorescence detection.
Assuming a thermal distribution of number states, the $\bar{n}$ can be estimated based on the sideband ratio averaged over a large number of outcomes \cite{leibfried2002trapped}.

The results for M1 are shown in Fig.~\ref{fig:fig3}\textbf{b}. With $N=1$, $\Tilde{p}_{0}$ (orange bar) =\,$p(\{d\}_{N})$=\,0.960(3), which differs from the initial population $p_{0}$ by 0.02. We attribute this discrepancy to detection error, mainly due to spin decoherence during the mapping sequence. 
For $N=2$ and $N=3$, $\Tilde{p}_{0}$ is very close to $p_0$ because the state is heralded multiple times, largely suppressing erroneous state declarations. However, we discard an increasing fraction of trials (7.8\% and 10.8\%, respectively) in post-selection when outcomes from the multiple rounds disagree. The discarded fraction is larger than that expected due to detection error alone, indicating that the motional state is changing slightly with increasing $N$, likely due to heating.
The results with mapping M2 are displayed in Fig.~\ref{fig:fig3}\textbf{c}. Here $\Tilde{p}_{0}$ =\,$p(\{b\}_{N})$, and thousands of photons are scattered with each detection of $\left | 0 \right \rangle$. The relative frequencies $\Tilde{p}_{0}$ also converge to the initial state population $p_0$, but more trials are discarded in post-selection, potentially due to imperfect protection of the Stretch mode from Mg$^{+}$ scattering. We discard 12.5\% ($N$=2) and 20.2\% ($N$=3) of all interrogations for M2. For larger $N$, heating during longer sequences increases leakage into higher number states which will lower the readout fidelity.
 
The $\bar{n}$ values of final motional states as determined by MAS/MSS transition probabilities are shown in Fig.~\ref{fig:fig3}\textbf{d}. The black data points show $\bar{n}$ when the motional state is left in the Alternating mode for a duration equivalent to running $N$ rounds of the measurement protocol; the increase in $\bar{n}$ is just due to heating of the mode. Red circles and red diamonds represent trials where the measurement blocks are executed with M1 and M2 respectively. The $\bar{n}$ values are reduced compared to just a delay because the motional state resides some of the time in the Stretch mode, which has a lower heating rate. Blue circles and blue diamonds show $\bar{n}$ for trials post-selected on the measurement heralding $\left|0\right\rangle$ $N$ times for M1 and M2. In all cases, M1 yields the lowest $\bar{n}$, but post-selecting on M2, despite the large number of scattered photons, also yields reduced $\bar{n}$ compared to no post-selection. The difference between M1 and M2 may arise from residual recoil heating of the Stretch mode, possibly from a non-zero $z^3$ contribution to the trapping potential (see Supplementary Material). Nevertheless, the Stretch mode is still largely protected from recoil during readout, and remains close to the state it was projected into during interrogation.
Tables and plots of the complete data sets for all $N$ and all measurement outcomes can be found in the Supplementary Material.

Coherent coupling of normal modes of a mixed-species ion string can be used for cooling~\cite{Modecouplingforcooling}, indirect state preparation~\cite{chou2017preparation}, and precision spectroscopy based on quantum logic~\cite{schmidt2005spectroscopy}.  
Generating suitable spatial variation and strength for the couplings in larger ion crystals can be challenging for some mode pairs. This is generally improved by using smaller traps with more control electrodes. If direct coupling of two modes is challenging, one can couple several pairs of modes sequentially for operations such as state transfer. 
Our approach can be combined with spin-motion control techniques to enable new quantum simulations~\cite{hartmann2006strongly,greentree2006quantum}. 
The non-destructive measurement protocol using protected modes can be adapted to measure any single bit of information about the motional state with an appropriate mapping sequence. This can be exploited for bosonic QEC codes~\cite{chuang1997bosonic,gottesman2001encoding, leghtas2013hardware, michael2016new,gottesman2001encoding,fluhmann2019encoding},
and for other applications requiring repeated motional state measurement. Symmetric strings with $2N+1$ ions have $N$ protected modes that could be used in demonstrations of multi-mode entangled bosonic states. 

We note that similar work on protected modes of trapped ion crystals is underway in other research groups~\cite{Metzner2021}. 

\bibliography{references.bib}

\begin{acknowledgments}
	We thank Hannah Knaack and Jules Stuart for helpful comments on the manuscript. P.-Y.H., J.J.W., S.D.E., G.Z. acknowledge support from the Professional Research Experience Program (PREP) operated jointly by NIST and the University of Colorado. 
	S.D.E. acknowledges support from the a National Science Foundation Graduate Research Fellowship under grant DGE 1650115. 
	D.C.C. and A.D.B. acknowledge support from a National Research Council postdoctoral fellowship.
	This work was supported by IARPA and the NIST Quantum Information Program.
\end{acknowledgments}

\pagebreak
\widetext

\begin{center}
\textbf{\large Methods}
\end{center}
\section{Coupling Hamiltonian derivation}\label{CHD}
\noindent We consider a linear string consisting of $N$ ions, with possibly different masses $M_{n}$ and charges $Q_{n}$ for $n\in\{1,...,N\}$, trapped in a three-dimensional potential well $U_{0}(\textbf{r})$ formed by externally applied potentials. We choose a coordinate system $(x,y,z)$ that is aligned with the principal axes of the equipotential ellipsoids that characterize $U_{0}(\textbf{r})$ near its minimum position, which we define as the origin of the coordinate system. As shown in Fig.~\ref{fig:fig1}\textbf{b}, $x$ and $y$ point from the ion positions toward the control and rf electrodes, and $z$ is along the trap axis which runs parallel to the electrode edges.
The coordinate origin is in the plane parallel to and midway between the electrode wafers and coincides with the minimum of the harmonic potential (black line) sketched in Fig.~\ref{fig:fig1}\textbf{c}.  The coordinate axes line up with the eigenvectors of three groups of normal (decoupled) motional modes, with $N$ modes in each group that we will derive next. The total potential energy for $N$ ions in the potential well at positions $\textbf{r}_n = (r_{x,n},r_{y,n},r_{z,n})^T$ is given by
\begin{equation}\label{Eqn:pot}
U_\mathrm{pot}(\textbf{r}_1,...\textbf{r}_N) = \sum_{n=1}^{N}Q_{n}U_{0}(\textbf{r}_{n}) + \sum_{n=1} ^{N}\sum_{n'>n} ^{N} \frac{Q_{n}Q_{n^{'}}}{4\pi\epsilon_{0}|\textbf{r}_{n}-\textbf{r}_{n^{'}}|}.    
\end{equation}
By simultaneously solving $\partial U_\mathrm{pot}/\partial \textbf{r}_{n}=0$ for all $n$, we obtain each ion's equilibrium position $\textbf{r}_{n}^{(0)}$. Expanding $U_\mathrm{pot}$ to second order in small, mass-weighted coordinate changes $q_{i,n}= (r_{i,n}-r_{i,n}^{(0)})/\sqrt{M_{n}}$ with  $i\in\{x,y,z\}$ around $\textbf{r}_{n}^{(0)}$ and diagonalizing the resulting Hessian matrix, we obtain $3 N$ mutually decoupled normal modes of ion motion with frequencies $\omega_{i,k}$ and quantized normal mode coordinates 
$$
u_{i,k}=\sqrt{\frac{\hbar}{2 \omega_{i,k}}}\left(\hat{a}_{i,k}+\hat{a}^\dag_{i,k}\right),
$$
where $\omega_{i,k}$, $\hat{a}^{\dagger}_{i,k}$, and $\hat{a}_{i,k}$ are the motional frequency, creation operator, and annihilation operator respectively of the $k$-th mode along axis $i\,\in\, \{x,y,z\}$. In the normal mode coordinates, the Hamiltonian of the motion of an ion string consists of $3 N$ uncoupled HOs and can be written as
$$
H_{0} = \sum_{i\in\{x,y,z\}}\sum_{k=1}^{N}\hbar \omega_{i,k}\left(\hat{a}^{\dagger}_{i,k}\hat{a}_{i,k}+1/2\right).
$$ 
Each ion oscillates around its equilibrium position, but does not participate in all normal modes equally in general and may not participate at all in some modes. For the $n$th ion the displacement along the  $i$-th axis, $\hat{q}_{i,n}$ can be written in terms of the $k$-th normal mode creation and annihilation operators as
\begin{equation}\label{Eqn:iontomode}
\hat{q}_{i,n} = \sum^{N}_{k=1}\sqrt{\frac{\hbar}{2M_{n}\omega_{i,k}}}\xi^{(i)}_{n,k}\left(\hat{a}_{i,k}+\hat{a}_{i,k}^{\dagger}\right),
\end{equation}
where $\xi^{(i)}_{n,k}$ is the transformation matrix element between the spatial coordinates of the $n$th ion displacement $q_{i,n}$ along axis $i$ and the normal mode vector component along the same axis for the $k$-th eigenmode.\\ 
\\
To couple two particular normal modes, mode $a$ oscillating at frequency $\omega_{i_{a},a}$ along axis $i_{a}$ and mode $b$ at frequency $\omega_{i_{b},b}$ along axis $i_{b}$, we can apply an oscillating perturbing potential $U_\mathrm{mod}(\textbf{r},t)=U_m(\textbf{r})\cos(\omega t+\phi)$ with $\omega$ close to the frequency difference of the modes we would like to couple, $\omega \approx \omega_{i_{a},a} - \omega_{i_{b},b}$. Expanding $U_m(\textbf{r})$ up to second order around a certain position $\textbf{r}_0$, we obtain 
\begin{equation}
U_m(\textbf{r}_0+\delta\textbf{r}) \approx U_m(\textbf{r}_0) + \sum_{i\in\{x,y,z\}} \frac{\partial U_m}{\partial i}|_{\boldsymbol{r}=\boldsymbol{r}_{0}} \delta r_{i} + \frac{1}{2}\sum_{i,j\in\{x,y,z\}} \frac{\partial^{2} U_m}{\partial i \partial j}|_{\textbf{r}=\textbf{r}_0}\delta r_{i} \delta r_{j}.
\end{equation}
Anticipating that only terms proportional to $\delta r_{i_{a}}\delta r_{i_{b}}$ of the two normal modes we desire to couple will rotate slowly in the interaction picture with respect to $H_{0}$, we can drop all other terms in the expansion of $U_m(\textbf{r})$:
$$
U_m(\textbf{r}_0+\delta\textbf{r}) \approx 2^{-\delta(i_a,i_b)} \frac{\partial^{2} U_m}{\partial i_a \partial i_b}|_{\textbf{r}=\textbf{r}_0}\delta r_{i_a} \delta r_{i_b}.
$$
Here $\delta(i_a,i_b) = 1$ for $i_a=i_b$ and 0 otherwise and we have used $\partial^{2} U_m /(\partial i_{a} \partial i_{b})=  \partial^{2} U_m /(\partial i_{b} \partial i_{a})$.
In practice, the dropped terms may cause undesirable distortion of the potential and excess ion motion and should be minimized when designing the perturbing potential. Again only keeping near-resonant terms, inserting the displacement operators for displacements of the $n$th ion in modes $a,b$, namely $\delta r_{i_{a},n}=\hat{q}_{i_{a},n}, \delta r_{i_{b},n}=\hat{q}_{i_{b},n}$, abbreviating ${\partial ^{2} U_m/(\partial i_{a}\partial i_{b}) |_{\boldsymbol{r}=\boldsymbol{r^{(0)}_{n}}} \equiv \alpha_{n}}$ and inserting Eq.\,(\ref{Eqn:iontomode}), the Hamiltonian from the perturbing potential can be approximated as 
\begin{equation}\label{Eqn:Up}
\begin{split}
H =& \sum_{n=1}^{N} Q_{n}U_{\rm mod}(\textbf{r}_n, t)\\
\approx& \sum_{n=1}^{N} Q_{n} 2^{-\delta(i_{a},i_{b})}\alpha_{n}\hat{q}_{i_{a},n}\hat{q}_{i_{b},n}\cos(\omega t+\phi)\\
    =& \sum_{n=1}^{N} Q_{n} 2^{-\delta(i_{a},i_{b})}\alpha_{n} \left[\sum_{k=1}^{N} \sqrt{\frac{\hbar}{2M_{n}\omega_{i_{a},k}}} \xi^{(i_{a})}_{n,k}\left(\hat{a}_{i_{a},k}+\hat{a}_{i_{a},k}^{\dagger}\right)\right]\\
    &\qquad\times\left[\sum_{l=1}^{N} \sqrt{\frac{\hbar}{2M_{n}\omega_{i_{b},l}}} \xi^{(i_{b})}_{n,l}\left(\hat{a}_{i_{b},l}+\hat{a}_{i_{b},l}^{\dagger}\right)\right]\\
    &\qquad\times\frac{1}{2}\left(e^{-i(\omega t+\phi)}+e^{i(\omega t+\phi)}\right)\\
    =& \sum_{n,k,l=1}^{N} 2^{-\delta(i_{a},i_{b})}\frac{\hbar Q_{n}\alpha_{n}}{4M_{n} \sqrt{\omega_{i_{a},k}\omega_{i_{b},l}}} \xi^{(i_{a})}_{n,k}\xi^{(i_{b})}_{n,l}\left(\hat{a}_{i_{a},k}+\hat{a}_{i_{a},k}^{\dagger}\right)
    \left(\hat{a}_{i_{b},l}+\hat{a}_{i_{b},l}^{\dagger}\right)\left(e^{-i(\omega t+\phi)}+e^{i(\omega t+\phi)}\right).
\end{split}
\end{equation}
\\
\\
We analyze this expression in the interaction frame with respect to $H_{0}$ by replacing $\hat{a}_{i,k}\rightarrow \hat{a}_{i,k} e^{-i\omega_{i,k}t}$, $\hat{a}_{i,k}^{\dagger}\rightarrow \hat{a}_{i,k}^{\dagger}e^{i\omega_{i,k}t}$. When $\omega = \omega_{i_a,a}-\omega_{i_b,b}$, we can neglect all terms that are not rotating at $\pm\left[\omega-\left(\omega_{i_a,a}-\omega_{i_b,b}\right)\right]$, which simplifies the coupling Hamiltonian\,(\ref{Eqn:Up}) to
\begin{equation}\label{Eq:CoupHMet}
H = \hbar g_0\left(e^{i\phi}\hat{a}\hat{b}^{\dagger} + e^{-i\phi} \hat{a}^{\dagger}\hat{b}\right)
\end{equation}
where we use $\omega_{i_a,a} =\omega_a$,  $\omega_{i_b,b} =\omega_b$,  $\hat{a}_{i_a,a} = \hat{a}$ and  $\hat{a}_{i_b,b}=\hat{b}$ for simplicity from this point onward. Note that coupling two modes along the same axis, $i_{a} = i_{b}$, results in two near-resonant cross-terms proportional to $\xi^{(i_{a})}_{n,a}\xi^{(i_{a})}_{n,b}$ and $\xi^{(i_{a})}_{n,b}\xi^{(i_{a})}_{n,a}$ that both contribute to the coupling equally and cancel the factor $2^{-\delta(i_{a},i_{a})}$. The coupling strength is 
\begin{equation}
g_0 = \sum_{n=1}^{N}g_{n} =\sum_{n=1}^{N} \frac{Q_{n}\alpha_{n}}{4M_{n}\sqrt{\omega_{a}\omega_{b}}}   \xi^{(i_{a})}_{n,a}\xi^{(i_{b})}_{n,b}.    
\end{equation}
This is identical to Eq.\,(\ref{Eq:CoupRat}) in the main text. 
\\
\\
\section{Time evolution of coupled motional states }\label{Sec:phaseshift} 
\noindent When two modes represented by ladder operators $\hat{a}$ and $\hat{b}$ are coupled by the Hamiltonian Eq.~(\ref{Eq:CoupHMet}), their states of motion will become entangled and, after an exchange of population, disentangled, in a periodic fashion. The time-dependent states can be found by first performing a basis transformation
\begin{eqnarray}
\hat{c}_{+} &=& \frac{1}{\sqrt{2}}\left(\hat{a}+e^{-i \phi}\hat{b}\right) \nonumber\\ 
\hat{c}_{-} &=& \frac{1}{\sqrt{2}}\left(\hat{a}-e^{-i \phi }\hat{b}\right),
\end{eqnarray}
which diagonalizes the interaction Hamiltonian 
\begin{equation}
\hbar g_0\left(e^{i\phi}\hat{a}\hat{b}^{\dagger} + e^{-i\phi} \hat{a}^{\dagger}\hat{b}\right) = \hbar g_0\left(\hat{c}_{+}^{\dagger}\hat{c}_{+} - \hat{c}_{-}^{\dagger}\hat{c}_{-}\right).  
\end{equation}
The right hand side represents two harmonic oscillators with energies separated by twice the interaction energy $\hbar g_0$. In the interaction frame of reference, these oscillators have simple equations of motion
\begin{equation}
\hat{c}^\dag_{\pm}(t)=\hat{c}^\dag_{\pm}(0)\exp(\pm i g_0 t).   
\end{equation}
Writing $\hat{a}^{\dag}(0)=\hat{a}^{\dag}$, $\hat{b}^{\dag}(0)=\hat{b}^{\dag}$  for brevity and inserting the time dependence into the equations for $\hat{a}(t)$ and $\hat{b}(t)$ yields 
\begin{eqnarray}\label{Eq:CouMod}
\hat{a}^\dag(t) &=& \hat{a}^\dag \cos(g_0 t)+ i e^{i \phi} \hat{b}^\dag \sin(g_0 t) \nonumber\\
\hat{b}^\dag(t) &=& \hat{b}^\dag \cos(g_0 t)+ i e^{-i \phi} \hat{a}^\dag \sin(g_0 t).
\end{eqnarray}
Any state of the oscillators at time $t$ can be written as a superposition of number states with complex amplitudes $c_{mn}$ by acting with different combinations of creation operators on the vacuum state $\left|0\right\rangle_a \left|0\right\rangle_b$,
\begin{equation}
\left|\Psi_a(t)\right\rangle\left|\Phi_b(t)\right\rangle= \sum_{m,n=0}^\infty \frac{c_{m n}}{\sqrt{m! n!}}\left[\hat{a}^\dag(t)\right]^m \left[\hat{b}^\dag(t)\right]^n \left|0\right\rangle_a\left|0\right\rangle_b,
\end{equation}
such that the time dependence is fully captured in the creation operators. For general times $t$ this implies a rather complicated entangled state of the modes, which becomes simpler for certain evolution times. For example when setting $\tau_{\rm BS}=\pi/(4 g_0)$ the trigonometric factors $\sin(g_0 \tau_{\rm BS})=\cos(g_0 \tau_{\rm BS})=1/\sqrt{2}$ and Eq.(\ref{Eq:CouMod}) turns into a beamsplitter relation~\cite{caves1981quantum} that can be used to demonstrate the Hong-Ou-Mandel effect, here for two modes at different frequencies in a mixed-species string of ions (See the main text and Fig.~\ref{fig:fig2}\textbf{f} and \ref{fig:fig2}\textbf{g}).\\
\\
Eq.(\ref{Eq:CouMod}) simplifies even more for $\tau_k = k \pi/(2 g_0)$ with $k$ a positive integer. For $k$ odd this yields
\begin{eqnarray}
\hat{a}^\dag(\tau_k) &=& i e^{i(g_0 \tau_k+ \phi)} \hat{b}^\dag\nonumber\\
\hat{b}^\dag(\tau_k) &=& i e^{i(g_0 \tau_k -\phi)} \hat{a}^\dag,
\end{eqnarray}
which implies that $\left|\Psi_a(\tau_k)\right\rangle\left|\Phi_b(\tau_k)\right\rangle$ with $k$ odd has the original states of modes $a$ and $b$ {\it swapped} and shifted by a phase $g_0 \tau_k$ per phonon, plus or minus $\phi$. This phase difference arises relative to that of the uncoupled evolution of the modes and can be thought of as a consequence of the coupling that modifies the energies of the eigenstates with the additional factors due to the phase $\phi$ of the applied drive. For $k$ even  
\begin{eqnarray}\label{Eq:evenkshift}
\hat{a}^\dag(\tau_k) &=& e^{i g_0 \tau_k}\hat{a}^\dag  \nonumber\\
\hat{b}^\dag(\tau_k) &=& e^{i g_0 \tau_k}\hat{b}^\dag,
\end{eqnarray}
which signifies one or several complete forth-and-back exchanges and a phase shift due to the coupling energy. Up to this phase shift, the state $\left|\Psi_a(\tau_k)\right\rangle\left|\Phi_b(\tau_k)\right\rangle$ with $k$ even is identical to the one at $t=0$ in the interaction frame of reference.   
\section{Coupling drive generation and control}

We use a segmented linear Paul trap consisting of a pair of RF electrodes and 47 control electrodes~\cite{blakestad2010transport}. The voltages of the control electrodes are produced by 47 independent arbitrary waveform generators (AWGs) with 16-bit resolution running at 50 MSPS (megasamples per second)~\cite{bowler2013arbitrary}. Each AWG output is connected to a control electrode through a two-stage low-pass filter with a 3\,dB corner frequency of about 50 kHz to suppress noise at motional frequencies. The oscillating potential $U(\textbf{r}, t)$ for creating mode-mode coupling is produced by applying suitable voltages to the twelve electrodes nearest to the ions using the corresponding AWG channels. The oscillating signals are added to the static voltages that produce the axial confinement. The AWGs are not actively synchronized, but have approximately equal clock speeds, so we reset their phase at the beginning of each experiment to make sure all the drives oscillate in phase. Coupling of motional modes becomes ineffective for  motional frequency differences larger than 1\,MHz due to attenuation from the low-pass filters and the 1 MHz bandwidth of the AWG output amplifiers.
\\
\\
We shape the amplitude envelope of coupling pulses to suppress the off-resonant excitation of other normal modes due to spectral side-lobes of the modulation pulses. The pulse amplitude ramps up as approximately $\sin^2(2\pi f t)$, with $f$= 12.5 kHz and 0\,$\leq t \leq$\,20\,$\mu$s, at the beginning of the pulse and ramps back to zero using a time-reversed copy of the ramp-up. We observe significant off-resonant excitation of the axial In-phase mode (at $\sim$\,1.5\,MHz) of a $^{9}$Be$^{+}$-$^{25}$Mg$^{+}$-$^{9}$Be$^{+}$ crystal when using a square coupling pulse near the resonant frequency of the Alternating-Stretch coupling, while such excitation is largely suppressed with shaped pulses.
\\
\\
We determine the coupling drive amplitudes for the twelve electrodes using a trap potential simulation \cite{blakestad2010transport}. The potential of each electrode is modeled by using the boundary element method~\cite{sadiku2000numerical}. We calculate the total potential around the center of an ion crystal by summing up the potential from all twelve electrodes.
We optimize the voltages to generate a potential for which the desired spatial derivative is maximized while the unwanted components are minimized. These unwanted terms typically include the gradients $\partial U_m /\partial i$, $i\in\{x,y,z\}$, which displace and potentially heat the ion motion, and the curvatures $\partial^{2} U_m /\partial i^{2}$, $i\in\{x,y,z\}$ which modulate motional frequencies. The undesired curvatures cannot be eliminated due to the constraint from the Laplace equation $\nabla^2U_m=0$, but they can be chosen to be far off-resonant. Higher-order derivatives of the potential are typically negligible in our trap and are not considered in the simulations.
\\
\\
\newpage
\widetext
\setcounter{equation}{0}
\renewcommand{\theequation}{S\arabic{equation}}
\setcounter{figure}{0}
\renewcommand{\thefigure}{S\arabic{figure}}
\setcounter{table}{0}
\renewcommand{\thetable}{S\arabic{table}}

\begin{center}
\textbf{\large Supplementary Material: Coherent coupling and non-destructive measurement of trapped-ion mechanical oscillators}
\end{center}

\section{Ion species and state manipulation}
\begin{figure}[h]
    \centerline{\includegraphics[width=1\textwidth]{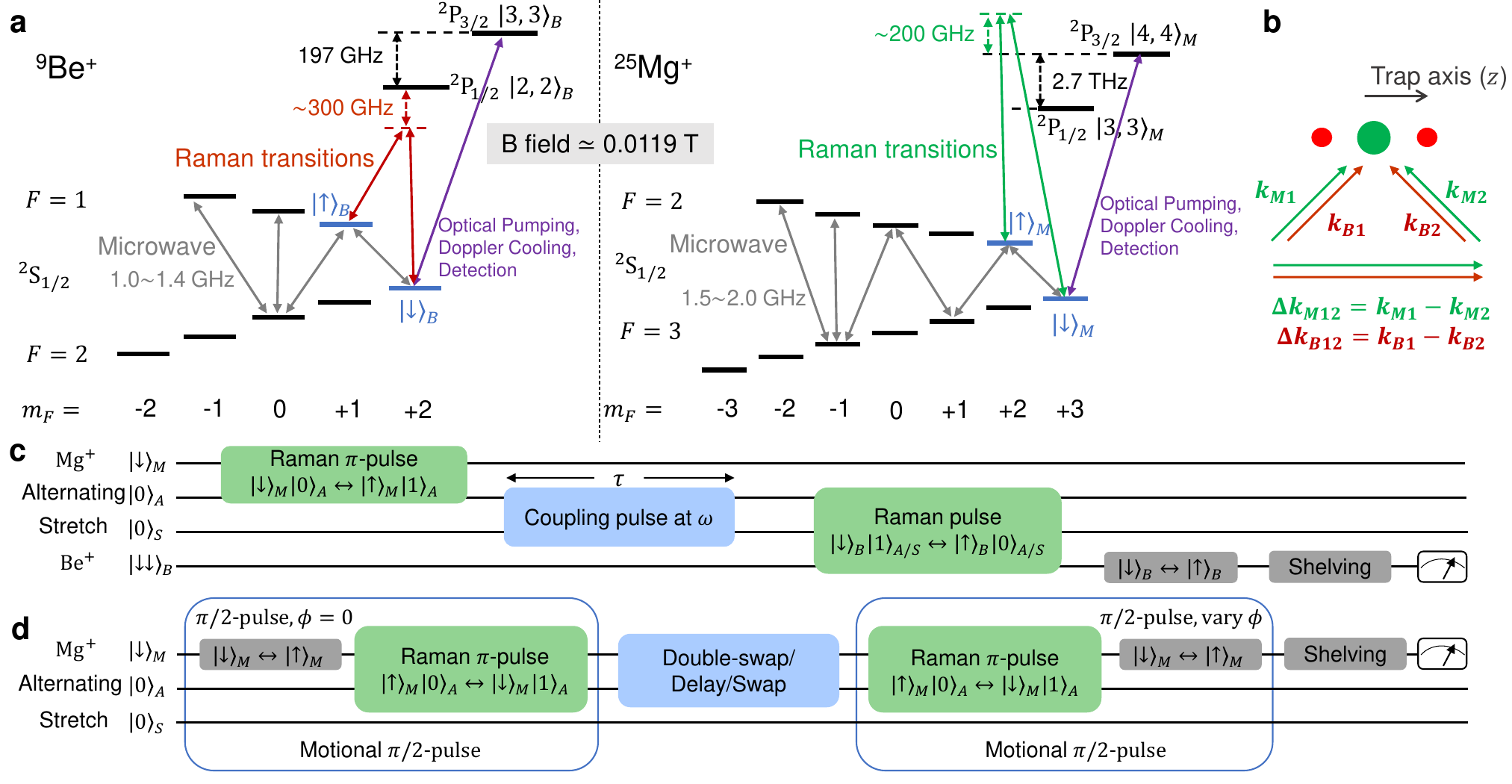}}
    \caption{\textbf{a}, Relevant electronic states of $^{9}$Be$^{+}$ and $^{25}$Mg$^{+}$ at an applied magnetic quantization field of $|\vec{B}|\approx$\,0.0119 T. \textbf{b}, Configuration of Raman laser beams for Be$^{+}$ (red) and Mg$^{+}$ (green). \textbf{c}, Experimental sequence for calibrating coupling resonant frequency and rates (results in Fig.~\ref{fig:fig2}\textbf{b,c}) of the Alternating-Stretch coupling.
    \textbf{d}, Experimental sequence for a motional Ramsey interference experiment between $\left|0\right\rangle_{A}$ and $\left|1\right\rangle_{A}$ (results in Fig.~\ref{fig:fig6}\textbf{c}) with a delay, or a coupling pulse for a single swap or a double swap inserted between the motional $\pi/2$ pulses.
     }
    \label{fig:fig4}
\end{figure}

\noindent We trap two ion species, $^{9}$Be$^{+}$ and $^{25}$Mg$^{+}$, at a quantization magnetic field of $\sim$\,0.0119\,T. The relevant electronic states of both species are illustrated in Fig.~\ref{fig:fig4}\textbf{a}.
A $\sigma^{+}$-polarized ultraviolet (UV) laser beam near 313\,nm optically pumps Be$^{+}$ ions to $\left| \downarrow\right\rangle_{B}=$ $^{2}S_{1/2}\left|F=2,m_F=2 \right \rangle_{B}$, while Doppler cooling and state-dependent fluorescence detection are implemented with a second UV laser beam driving the $ ^{2}S_{1/2}\left|2,2 \right \rangle_{B} \leftrightarrow \,^{2}P_{3/2}\left|3,3 \right\rangle_{B}$ cycling transition, causing photons to be emitted from the ion when Be$^{+}$ is in the ``bright" state $\left| \downarrow\right\rangle_{B}$.
Similarly, Mg$^{+}$ ions are optically pumped to the bright state $\left| \downarrow\right\rangle_{M}$= $^{2}S_{1/2}\left|3,3 \right \rangle_{M}$ with a $\sigma^{+}$-polarized laser beam near 280\,nm. A second UV laser beam is used to drive the $^{2}S_{1/2}\left|3,3 \right \rangle_{M} \leftrightarrow \,^{2}P_{3/2}\left|4,4 \right\rangle_{M}$ transition for Doppler cooling and fluorescence detection. We denote $\left|\uparrow\right\rangle_B =\,^{2}S_{1/2} \left|F=1,m_F=1\right\rangle_B$ and $\left|\uparrow\right\rangle_M =\, ^{2}S_{1/2} \left|F=2,m_F=2\right\rangle_M$. 
State-dependent fluorescence detection is accomplished with resonant UV light illuminating the ions for a duration of 330\,$\mu$s for Be$^{+}$ and 200\,$\mu$s for Mg$^{+}$, with a fraction of the ion fluorescence collected by an achromatic imaging system and detected by a photomultiplier tube. 
To distinguish two hyperfine states of interest, we apply a ``shelving" sequence that consists of microwave $\pi$ pulses to transfer one hyperfine state to the bright state  and the other to a dark state (a hyperfine state away from the bright state in the $^{2}S_{1/2}$ manifold) before fluorescence detection. The microwave transitions used in the shelving sequence are indicated with grey arrows in Fig.~\ref{fig:fig4}\textbf{a}. The detected photon counts approximately follow Poisson distributions with a mean of $\sim$\,30 counts per detection for each ion when they are in the bright states $\left|\downarrow\right \rangle_{B}$ and $\left|\downarrow\right \rangle_{M}$. The ions scatter only a few photons per detection when in any other hyperfine states. In particular, we detect a background of $\sim$\,2 photons per detection for Be$^{+}$ and $\sim$\,1 photons for Mg$^{+}$ (dominated by background scatter) when ions are in the ``dark" states, $^{2}S_{1/2}\left|1,-1\right\rangle_{B}$ and $^{2}S_{1/2}\left|2,-2\right\rangle_{M}$.\\
\\
In the coupling calibrations, we analyze the photon counts of a reference dataset by using maximum likelihood estimation (MLE) to determine the Poissonian mean photon counts of $N$=\,0,1,2 Be$^{+}$ ions in the bright state. Then, we determine the probability $P_{b}(N)$ of $N$ Be$^{+}$ ions in the bright state for the coupling calibration data using MLE with the fixed and pre-determined Poissonian means.
Correlations of the populations in different motional modes need to be evaluated within a single experimental trial. We choose photon count thresholds such that the number of bright ions is distinguished with minimal error for both ion species in each trial and obtain the probability of the joint state over many experiment repetitions. For two Be$^{+}$ ions, the ions are identified to be both in the bright state when the counts $c_{Be} > 46$; and only one bright Be$^{+}$ ion when $13 \leq c_{Be} < 46$; zero bright ions otherwise. A single Mg$^{+}$ ion is determined to be in the bright state when the photon count $c_{Mg} >$\,9, and in the dark state otherwise. The histogram for one bright ion has an overlap of about 1.4\% with that of two bright ions and has nearly zero overlap with that of zero bright ions assuming ideal Poisson distributions. In the repeated motional state measurements, the single Mg$^{+}$ ion population is determined by using the threshold method.
\\
\\
We employ stimulated Raman transitions with two laser beams that coherently manipulate the internal states of the ions and the normal modes of the ion string. The Raman beams together with resonant repumping light are used to sideband-cool motional modes close to their ground states, and the Raman beams are used to prepare initial motional states and map the final motional states onto internal states of ions for readout. As illustrated in Fig.~\ref{fig:fig4}\textbf{b}, two pairs of Raman beams, one pair for Be$^{+}$ near 313\,nm (red arrows) and the other pair for Mg$^{+}$ near 280\,nm (green arrows), have their wave vector difference aligned with the $z$ axis, such that sideband transitions only address axial modes. The Alternating mode and the Stretch mode are cooled to an average quantum number of $\bar{n}\approx$ 0.07 and 0.02 respectively, while the third axial normal mode, the in-phase mode (at $\sim 2\pi\times$1.5 MHz), is cooled to a higher $\bar{n}\approx$ 0.25, because the cooling competes with a larger heating rate of $\sim$750 quanta per second in this mode. We measure the heating rates of the Alternating mode and the Stretch mode to be $\sim$\,60 and $\sim$\,1 quanta per second, respectively. In the repeated motional state measurements, the Alternating mode is cooled to a lower $\bar{n}\approx$ 0.02 than stated above, mainly due to increased Mg$^{+}$ Raman laser power compared to the other experiments.
\\
\\
When determining correlations between populations in different modes, we tailor the frequency and pulse shape of the Raman beams to realize sideband rapid adiabatic passage (RAP) pulses \cite{allen1975optical} that can implement nearly complete quantum state transfers $\left|\downarrow\right\rangle|n\rangle \leftrightarrow \left|\uparrow\right\rangle\left|n-1\right\rangle$ simultaneously for a range of $n$, despite sideband transitions having different $n$-dependent Rabi frequencies for different number states.
For example, the initial state $\left|\downarrow\downarrow\right\rangle_{B}\left|2\right\rangle_{S}$ of two Be$^{+}$ ions and the Stretch mode can be fully transferred to $\left|\uparrow\uparrow\right\rangle_{B}\left|0\right\rangle_{S}$ by an ideal RAP pulse, while a pulse with fixed frequency and square intensity envelope cannot transfer the full population between these two states. Similarly, the states $\left|\downarrow\right\rangle_{M}\left|n\right\rangle_{A}$ of a Mg$^{+}$ ion and the Alternating mode can be transferred to $\left|\uparrow\right\rangle_{M}\left|n-1\right\rangle_{A}$ simultaneously for all relevant $n>0$ with a single RAP pulse. 
To experimentally generate a RAP pulse, we shape the amplitude of two Raman beams to follow a truncated Gaussian envelope $A_{i} \exp(-t^{2}/t_{width}^{2})$ with $A_{i=1,2}$ the maximum pulse amplitudes of the two Raman beams at $t$=0 when the pulse amplitude is at peak, and $t_{width}$ sets the scale of the pulse duration. During this pulse, the relative detuning of one of the beams from the Raman resonance of the target transition is linearly swept from $-\delta_{max}$ to $\delta_{max}$ such that it is on resonance at $t=\,0$. The Gaussian envelope is truncated to zero at $\pm 2 \,t_{width}$.
The Be$^{+}$ RAP pulse uses $t_{width}$\,=\,400\,$\mu$s, $\delta_{max}$=\,\,0.25\,MHz for the Stretch mode sideband transition and the fidelity of single transfer is estimated to be $\sim$\,90\% through independent experiments. The Mg$^{+}$ RAP pulse on the Alternating mode sideband transition uses $t_{width}$\,=\,100\,$\mu$s, $\delta_{max}$\,=\,0.3\,MHz and the fidelity is estimated to be $\sim$\,96\%.
 
\section{Alternating-Stretch coupling characterization}
\noindent In order to characterize the Alternating-Stretch mode coupling (experimental sequence in Fig.~\ref{fig:fig4}\textbf{c}), we prepare both modes in the ground state and all three ions in their bright states $\left| \downarrow \right \rangle_{B/M}$. We create a single phonon in the Alternating mode with a $\pi$ pulse on the $\left| \downarrow \right\rangle_{M} \left|0\right\rangle_{A} \leftrightarrow \left| \uparrow \right \rangle_{M}\left|1\right\rangle_{A}$ MAS transition of the Mg$^{+}$ ion. 
Next, we apply a coupling pulse of variable frequency or duration to transfer the single phonon between modes. 
The probability of finding the single phonon in the Alternating mode $\,P$($n_{A}$=1) or the Stretch mode $P$($n_{S}$=1) varies as a function of coupling pulse frequency or duration. After the coupling pulse, we apply a MSS pulse on the $\left| \downarrow \right\rangle_{B} \left|n\right\rangle_{A/S} \leftrightarrow \left| \uparrow \right \rangle_{B}\left|n-1\right\rangle_{A/S}$ transition with a duration $t_{max}$, which is calibrated by finding the maximum probability of $\left|\uparrow \uparrow\right\rangle_{B}$ after applying a sideband pulse onto the two Be$^{+}$ ions prepared in $\left|\downarrow \downarrow\right\rangle_{B}$.
Before state-dependent fluorescence detection of Be$^{+}$, we exchange the population between $\left| \uparrow \right \rangle_{B}$ and $\left| \downarrow \right \rangle_{B}$ with a microwave $\pi$ pulse, then apply a shelving sequence to transfer the population in $\left| \uparrow \right \rangle_{B}$ to the dark state.
We obtain the probabilities $P_{b}(N)$ of $N$ bright Be$^{+}$ ions for $N\in\{0,1,2\}$ by using MLE as described above with the fluorescence histogram averaged over 300 experimental trials. We then use $P_{b}(N)$ to determine the populations of the three number states based on numerical simulation of the system. The numerical model assumes a square MSS pulse with a duration of $t_{max}$ that addresses both Be$^+$ ions with equal Rabi frequency and limits the state space of the Alternating and Stretch modes to the lowest three number states, $i.e. \sum_{n=0}^{2}P$($n_{A/S}$)=1. The model predicts $P$($n_{A/S}$=1)\,=\,$P_{b}$(1)/0.942, $P$($n_{A/S}$=2)\,=\,$P_{b}$(2)/0.889, and $P$($n_{A/S}$=0)\,=\,1-$P$($n_{A/S}$=1)-$P$($n_{A/S}$=2). For example, when the mode is in $|n=1\rangle$, the model predicts a probability of one of two ions flipping to be $P_b(1)=0.942$.
\\
\\
The results of a frequency scan are shown in Fig.~\ref{fig:fig2}\textbf{b} of the main text, where the coupling resonance is discernible from nearly complete exchange of probability from $P$($n_{A}$=1) to $P$($n_{S}$=1) around the frequency difference of the two modes, as expected. The data points from both modes are fitted to $P(\omega)= A \Omega_{0}^2\sin ^{2}(\Omega T/2)/\Omega^{2}+P_{0}$ with $\Omega=\sqrt{\Omega_{0}^{2}+(\omega-\omega_{0})^{2}}$ to yield a resonant frequency $\omega_{0}\approx 2\pi\times$\,0.283\,MHz.
When setting the coupling frequency at $\omega_{0}$ and scanning the coupling duration, we obtain the results shown in Fig.~\ref{fig:fig2}\textbf{c}, where two anti-correlated sinusoidal oscillations of $P$($n_{A}$=1) and $P$($n_{S}$=1) were observed and fit with $P(\tau)=A \sin(\Omega_{C}\tau+\phi_{c})\exp(-\tau/\tau_c)+y_{0}$. The data sets for both modes yield the same exchange rate $\Omega_{C}/(2\pi)\approx5.1$~kHz, $A\approx0.46$ and $y_{0}\approx0.47$. The coherence time is $\tau_{c,E}$=\,24(14)\,ms for $P$($n_{A}$=1) and $\tau_{c,S}$=\,19(11)\,ms for $P$($n_{S}$=1), both of which are approximately 200 times longer than the duration of a swap. 
The maximum $P$($n_{A/S}$=1)\,$\approx$\,0.93 deviates from the ideal value of 1, predominantly due to imperfect ground state cooling and imperfect single phonon injection into the Alternating mode.
\\
\section{Motional coherence after a coupling pulse}
To examine whether motional coherence is preserved after a coupling pulse, we perform a motional Ramsey-like experiment on the Alternating mode with the pulse sequence shown in Fig.~\ref{fig:fig4}\textbf{d}. We prepare both modes in the ground state and the Mg$^{+}$ ion in $\left|\downarrow\right\rangle_M \equiv \,^{2}S_{1/2} \left|F=3,m_F=3\right\rangle_M$ ($\left|\uparrow\right\rangle_M\equiv \,^{2}S_{1/2}\left|F=2,m_F=2\right\rangle_M$). 
Then, we apply an effective motional $\pi/2$ pulse consisting of a microwave carrier $\pi/2$ pulse on $\left|\downarrow\right\rangle_{M} \leftrightarrow \left|\uparrow\right\rangle_{M}$ and a subsequent sideband $\pi$ pulse on $\left|\uparrow\right\rangle_{M}\left|0\right\rangle_{A}\rightarrow \left|\downarrow\right\rangle_{M}\left|1\right\rangle_{A}$, which creates the superposition $\frac{1}{\sqrt{2}}(\left|0\right\rangle_{A}+\left|1\right\rangle_{A})$ in the Alternating mode and rotates the Mg$^{+}$ back to $\left|\downarrow\right\rangle_{M}$. Next, we either apply a double-swap operation (Double-swap), a single swap operation (Swap), or just a delay of the same duration as the double-swap pulse (Delay). 
Afterwards, a second sideband $\pi$ pulse transfers the superposition of number states back onto a superposition of Mg$^+$ internal states. A second microwave carrier $\pi/2$ pulse with phase difference $\phi$ with respect to the first $\pi/2$ pulse maps the motional phase difference between $\left|0\right\rangle_{A}$ and $\left|1\right\rangle_{A}$ onto the internal state populations of $\left|\downarrow\right\rangle_{M}$ and $\left|\uparrow\right\rangle_{M}$, which are then measured. 
In Fig.~\ref{fig:fig6}\textbf{c} of the main text, we show the data for the three cases discussed above, with fit lines to the function $P(\phi)=B\sin(\phi)+y_{0}$. The fit to the double-swap signal has a contrast (defined as $B/y_{0}$) of 0.95(1), higher than the contrast of 0.92(1) when performing a delay, indicating that the coupling drive causes no damage to the motional coherence but rather helps in preserving it longer, since the state is swapped into the Stretch mode where it experiences a lower heating rate. We also observe a phase shift of roughly $\pi$ between those two traces because a double-swap pulse not only exchanges the motional population between two modes back and forth, but also leads to number-state-dependent phase shifts as predicted by Eq.\,(\ref{Eq:evenkshift}). The rest of the contrast loss is mainly due to imperfections in state preparation and readout. 
When a single swap operation is performed, the state of the Alternating mode is replaced with the approximate Stretch mode ground state, removing the possibility for Ramsey interference and yielding roughly equal populations of both internal states, independent of the relative phase of the second $\pi/2$ pulse.
\\
\section{Joint motional population measurement} 
\begin{figure}
    \centering
    \centerline{\includegraphics[width=1\textwidth]{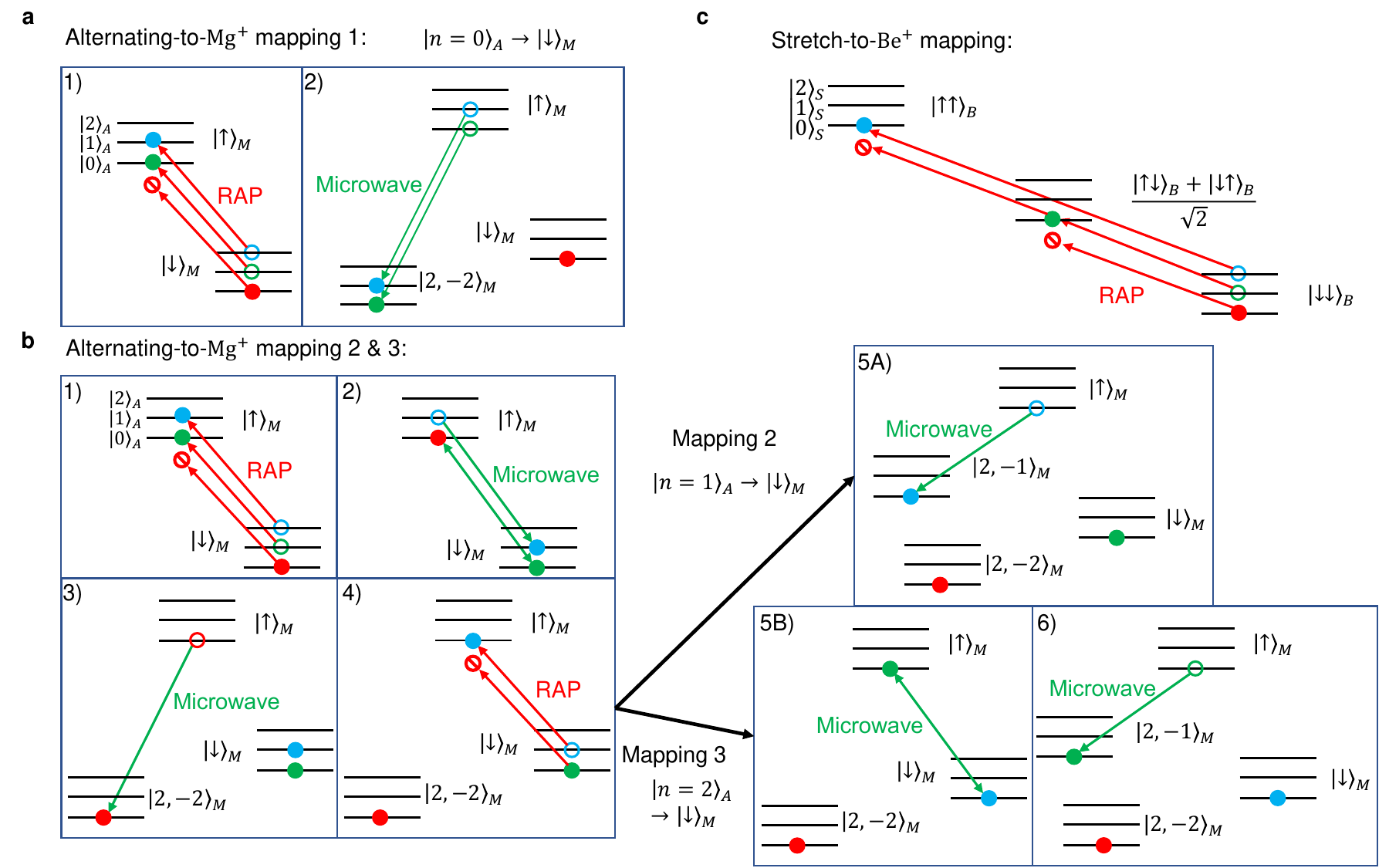}}
    \caption{Diagram of motion-to-spin mapping for determining joint populations of motional modes. Circles in different colors represent the initial population in $\left|n=0\right\rangle$ (red), $\left|1\right\rangle$ (green), $\left|2\right\rangle$ (blue) at the beginning of the illustrated step and solid dots at the end. Transitions are indicated by arrows and forbidden transitions are indicated by a prohibitory symbol at the tip of the arrow.}
    \label{fig:MapSeq}
\end{figure}

\begin{figure}[h]
    \centerline{\includegraphics[width=0.7\textwidth]{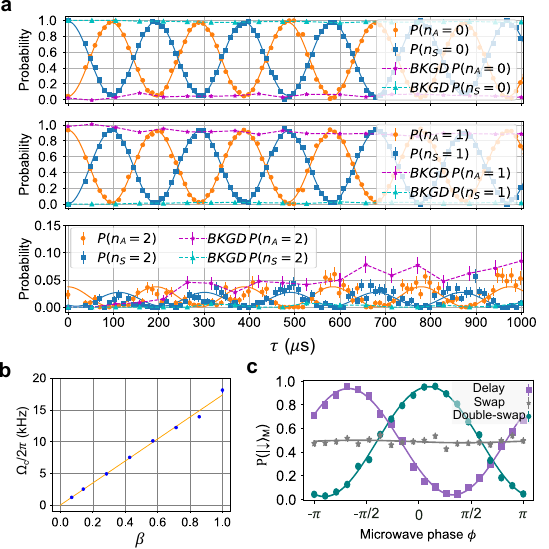}}
    \caption{\textbf{Characterization of the Alternating-Stretch coupling.} \textbf{a}, With the Alternating and Stretch mode prepared in $\left|1\right\rangle_{A}\left|0\right\rangle_{S}$, the probabilities of the Alternating (orange dots) and Stretch (blue squares) modes in number state $n$=0 (top), $n$=1 (middle, also partially shown in Fig.~\ref{fig:fig2}\textbf{c}), $n$=2 (bottom), oscillate as the coupling pulse duration is varied while the coupling frequency is held on resonance. As references, we repeat the experiment by replacing the coupling pulse with a delay of the same duration to investigate effects of heating, and show the corresponding probabilities of certain number states in the Alternating (magenta stars) and Stretch (cyan triangles) modes in the three panels. The population mostly stays in $n$=0 and $n$=1, while oscillating out of phase between the two modes at $\Omega_{c}=2g_{0}$. In the bottom panel, the populations in $n$=2 of the two modes show small oscillations (due to imperfect state preparation) on top of a slowly growing background that is roughly the average heating rate of the two coupled modes, as independently verified by the reference data (magenta stars and cyan triangles). Solid lines are fits to the data and dashed lines are the guides to the eye.
    \textbf{b}, The population exchange rate $\Omega_{c}$ can be varied by controlling the amplitude of the coupling potential with a relative factor $0\leq \beta \leq 1$, such that $U(\textbf{r}) = \beta U_{max}(\textbf{r})$. This is accomplished by scaling the oscillating potential amplitude of electrode $i$ with $V_{i} = \beta V_{max,i}$ for all twelve electrodes ($i=1,2,...,12$). We choose $\beta$=0.286 for all other results presented in this work. Using higher $\beta$ causes larger unintended excitation of the INPH mode.
    \textbf{c}, Motional coherence verification after a coupling pulse. A superposition of $\left|0\right\rangle_A$ and $\left|1\right\rangle_A$ undergoes a swap operation (grey stars), a double-swap operation (teal dots), or a delay (purple squares) equivalent to the duration of the double-swap operation. The resulting motional superposition is mapped onto Mg$^+$ internal states with variable phase $\phi$. 
     } 
    \label{fig:fig6}
\end{figure}

To uncover correlations between the populations of the Alternating and Stretch modes, one needs to  determine their joint populations within a single experiment. Individual addressing of the Mg$^{+}$ and Be$^{+}$ ions with distinct wavelengths for laser-driven operations allows the Alternating and Stretch mode states to be mapped onto and detected via internal states of the two species respectively.
\\
\\
We prepare four different joint number states $\left|0\right\rangle_{A}\left|0\right\rangle_{S}$, $\left|1\right\rangle_{A}\left|0\right\rangle_{S}$, $\left|1\right\rangle_{A}\left|1\right\rangle_{S}$, and $\left|0\right\rangle_{A}\left|2\right\rangle_{S}$. The internal states of all three ions are initialized (and reset) in the $\left| \downarrow\right \rangle_{B/M}$ state before (and after) motional state preparation. The details of the preparation of each state are as follows:
\begin{itemize}
\item $\left|0\right\rangle_{A}\left|0\right\rangle_{S}$ is prepared by sideband cooling all three axial modes close to their ground states, with an infidelity of 0.09, estimated by using the average occupations $\bar{n}$ of the Alternating and Stretch modes, determined from sideband ratio measurements.
\item $\left|1\right\rangle_{A}\left|0\right\rangle_{S}$ is prepared from $\left|0\right\rangle_{A}\left|0\right\rangle_{S}$ with a microwave $\pi$ pulse on the $\left| \downarrow \right\rangle_{M} \rightarrow \left| \uparrow \right\rangle_{M}$ transition, followed by a sideband $\pi$ pulse on $\left| \uparrow \right\rangle_{M}\left|0\right\rangle_{A} \rightarrow \left| \downarrow \right\rangle_{M}\left|1\right\rangle_{A}$, which injects a single phonon into the Alternating mode and leaves Mg$^{+}$ in $\left| \downarrow\right\rangle_{M}$. The microwave pulse has negligible error and the sideband $\pi$ pulse has an error of about 0.03, in large part due to the Debye-Waller effect from the axial in-phase mode~\cite{wineland1998experimental}. When such an error occurs, the sideband $\pi$ pulse is incomplete and leaves the Mg$^{+}$ ion partially in $\left| \uparrow\right\rangle_{M}$. Therefore, we apply an additional dissipative laser repumping pulse to ensure the Mg$^{+}$ ion is reset to $\left| \downarrow\right\rangle_{M}$ before joint state mapping.

\item $\left|1\right\rangle_{A}\left|1\right\rangle_{S}$ is prepared by initializing in $\left|1\right\rangle_{A}\left|0\right\rangle_{S}$, as described above. Then, a calibrated swap pulse transfers (with an error of approximately 0.01) the single phonon from the Alternating mode to the Stretch mode ($\left|0\right\rangle_{A}\left|1\right\rangle_{S}$). Then, another single phonon is injected into the Alternating mode using the same method described above, resulting in $\left|1\right\rangle_{A}\left|1\right\rangle_{S} \left| \downarrow\right\rangle_{M}$. 

\item $\left|0\right\rangle_{A}\left|2\right\rangle_{S}$ is prepared from $\left|0\right\rangle_{A}\left|0\right\rangle_{S}$ by injecting two phonons into the Stretch mode by globally addressing two Be$^{+}$ ions with a microwave carrier $\pi$ pulse on $\left| \downarrow\right\rangle_{B} \rightarrow \left| \uparrow \right\rangle_{B}$, followed by a sideband RAP pulse (with an error of approximately 0.05) on the transition $\left| \uparrow \right\rangle_{B}|n\rangle_{S} \rightarrow \left| \downarrow\right\rangle_{B}\left|n+1\right\rangle_{S}$. Afterwards, a Be$^{+}$ repumping pulse is applied to reset the internal states to $\left| \downarrow \downarrow\right\rangle_{B}$ in case the RAP pulse did not leave the ions in that state. 
\end{itemize}
For motional state analysis, we map the population of joint-number states in the subspace $\mathcal{S}\,=\,\{\left|0\right\rangle_{A},\left|1\right\rangle_{A},\left|2\right\rangle_{A}\}\bigotimes\{\left|0\right\rangle_{S},\left|1\right\rangle_{S},\left|2\right\rangle_{S}\}$ onto the internal states of two ion species. Two steps, Alternating-to-Mg$^{+}$ mapping and Stretch-to-Be$^{+}$ mapping, are sequentially implemented. The mapping is described in detail in the following and illustrated in Fig.~\ref{fig:MapSeq}.
\begin{itemize}
\item Alternating-to-Mg$^{+}$ mapping: One of three different mapping sequences (Fig.~\ref{fig:MapSeq}\textbf{a}, \textbf{b}) maps the population in one of the three lowest number states respectively onto the bright state $\left|\downarrow\right\rangle_{M}$ and shelves the other two number states to dark states, $\left|2, -1\right\rangle_{M}$ and $\left|2, -2\right\rangle_{M}$. Repeated experimental trials with different choices of mapping sequences are used to build statistics for the populations of $|0\rangle$, $|1\rangle$, and $|2\rangle$.
\\
\\
The mapping 1 for $\left|0\right\rangle_{A}\rightarrow\left|\downarrow\right\rangle_{M}$ (Fig.~\ref{fig:MapSeq}\textbf{a}) is as follows: 1) A MSS RAP pulse on the $\left|\downarrow\right\rangle_{M}|n\rangle_{A} \rightarrow \left|\uparrow\right\rangle_{M}\left|n-1\right\rangle_{A}$ transition flips the internal state of the Mg$^{+}$ ion from $\left|\downarrow\right\rangle_{M}$ to $\left|\uparrow\right\rangle_{M}$ if the Alternating mode is in $\left|1\right\rangle_{A}$ (green dot) or $\left|2\right\rangle_{A}$ (blue dot) while also subtracting one quantum of motion, but leaves $\left|\downarrow\right\rangle_{M}\left|0\right\rangle_{A}$  (red dot) unchanged. 2) A microwave pulse shelves the population in $\left|\uparrow\right\rangle_{M}$ (the initial population of $\left|1\right\rangle_{A}$ and $\left|2\right\rangle_{A}$) to $\left|2,-2\right\rangle_{M}$. 
\\
\\
The mapping 2 for $\left|1\right\rangle_{A}\rightarrow\left|\downarrow\right\rangle_{M}$ and the mapping 3 for $\left|2\right\rangle_{A}\rightarrow\left|\downarrow\right\rangle_{M}$ (Fig.~\ref{fig:MapSeq}\textbf{b}) start similarly. 1) A MSS RAP pulse separates the population of $\left|0\right\rangle_{A}$ from $\left|1\right\rangle_{A}$ and $\left|2\right\rangle_{A}$. 2) Then, a microwave $\pi$ pulse $\left|\uparrow\right\rangle_{M} \leftrightarrow \left|\downarrow\right\rangle_{M}$ is applied. 3) A microwave shelving sequence then transfers the population in $\left|\uparrow\right\rangle_{M}$ (the initial population of $\left|0\right\rangle_{A}$) to $\left|2,-2\right\rangle_{M}$. 4) A second MSS RAP pulse separates the initial population of $\left|1\right\rangle_{A}$ from that of $\left|2\right\rangle_{A}$. 
5A) The last step of mapping 2 is to apply a microwave shelving sequence which transfers the initial population of $\left|2\right\rangle_{A}$ (now in $\left| \uparrow \right \rangle_{M}$) to another dark state $\left|2,-1\right\rangle_{M}$ and leaves the initial population of $\left|1\right\rangle_{A}$ still in $\left|\downarrow\right\rangle_{M}$.
5B) For mapping 3, a microwave $\pi$ pulse of $\left|\uparrow\right\rangle_{M} \leftrightarrow \left|\downarrow\right\rangle_{M}$, inserted between the second RAP pulse and the final shelving sequence, maps the population in $\left|2\right\rangle_{A}$ onto $\left|\downarrow\right\rangle_{M}$ instead. 6) The initial population of $\left|1\right\rangle_{A}$ is shelved to $\left|2,-1\right\rangle_{M}$.
\item Stretch-to-Be$^{+}$ mapping (Fig.~\ref{fig:MapSeq}\textbf{c}) uses a RAP pulse on the $\left|\downarrow\right\rangle_{B}|n\rangle_{S} \rightarrow \left|\uparrow\right\rangle_{B}\left|n-1\right\rangle_{S}$ transition, flipping both Be$^{+}$ ions from $\left|\downarrow \right\rangle_{B}$ to $\left|\uparrow \right\rangle_{B}$ if $\left|n\right\rangle_{S}=\left|2\right\rangle$, which realizes $\left|\downarrow\downarrow \right\rangle_{B} \left|2\right\rangle_{S} \rightarrow \left|\uparrow\uparrow \right\rangle_{B} \left|0\right\rangle_{S}$; only one ion flips if $\left|n\right\rangle_{S}=\left|1\right\rangle$, $\left|\downarrow\downarrow \right\rangle_{B} \left|1\right\rangle_S \rightarrow 1/\sqrt{2}(\left|\uparrow\downarrow \right\rangle_{B} + \left|\downarrow\uparrow \right\rangle_{B}) \left|0\right\rangle_S$; and the two ions remain in $\left|\downarrow\downarrow \right\rangle_{B}$ when $\left|n\right\rangle_{S}=\left|0\right\rangle$. Before fluorescence detection, the population in $\left|\uparrow\right\rangle_{B}$ of both Be$^{+}$ ions is shelved to the dark state $\left|1,-1\right\rangle_{B}$ (not shown in Fig.~\ref{fig:MapSeq}\textbf{c}). 
\end{itemize} 
State-dependent fluorescence detection of Mg$^{+}$ and Be$^{+}$ is performed sequentially after the mapping steps described above to obtain photon counts of both species.
We perform three sets of experimental trials with different Alternating-to-Mg$^{+}$ mappings with $N$=1000 trials per set. In each experimental trial, the Be$^{+}$ counts $c_{Be}$ are compared to thresholds $\{$13, 46$\}$ to determine the number of the bright ions $N_{Be}$, which in turn indicates the Stretch mode state based on the mapping\\ \centerline{$c_{Be}>$\,46 $\rightarrow N_{Be}$\,=\,2 $\rightarrow n_{S}$\,=\,0; 13\,$< c_{Be}\leq$\,46 $\rightarrow N_{Be}$\,=\,1 $\rightarrow n_{S}$\,=\,1; $c_{Be}\leq$\,13 $\rightarrow N_{Be}$\,=\,0 $\rightarrow n_{S}$\,=\,2.}\\
When mapping $\left|n\right\rangle_{A}\rightarrow\left|\downarrow\right\rangle_{M}$, the Alternating mode is declared to be in $\left|n\right\rangle_{A}$ if the Mg$^{+}$ counts are beyond a threshold of nine. Experiments with the Mg$^{+}$ counts below the threshold of nine are discarded because the Alternating mode is likely in one of the other two number states and we cannot distinguish them in such cases. 
We calculate the populations of the nine joint states $\{|n\rangle_{A}|m\rangle_{S}\}$ ($n,m$ = 0,\,1,\,2) according to $P_{i}$=$N_{i}/N_{rep}$, ($i$=0,...,9) with $N_{i}$ the occurrence of the $i$-th joint state and $N_{rep}$ the total number of successful repetitions of all three mappings ($N_{rep}\approx N$). 
Uncertainties are calculated assuming that projection noise is the dominant noise source, $\Delta P_{i} = \sqrt{P_{i}(1-P_{i})/N_{rep}}$. 
\\
\\ 
Experimental results shown in Fig.~\ref{fig:fig2}\textbf{d}-\textbf{g} have appreciable state preparation and measurement (SPAM) errors. State preparation errors are from imperfect sideband cooling and imperfect phonon injection, as mentioned above. 
Measurement errors arise from imperfect mapping operations consisting of RAP pulses ($\sim$5\% error per RAP pulse) and microwave pulse sequences. Since the mapping process takes between 2.6\,ms and 3.8\,ms, heating during mapping can change the motional state, with the Alternating mode being more affected than the Stretch mode. The motional state measurement error from heating varies for different number states and scales approximately with $n_{A}+1/2$ since the Alternating mode heating rate is much higher than that of the Stretch mode. The readout error due to the threshold method is estimated to be negligible. Precise determination of how SPAM errors compound for each joint state is complicated, so we did not attempt this. The error sources discussed here approximately explain the SPAM errors observed in the experiments.\\
\\
\section{Beamsplitter fidelity lower bound}

We performed a 50:50 beamsplitter operation on two motional modes of a three-ion crystal.
Since the implemented beamsplitter is noisy, we wish to characterize its deviation from the ideal beamsplitter dynamics.
We quantify this deviation with the average fidelity $F$ of the beamsplitter on the one-phonon  subspace of the two modes.
In the subsection ``Derivation of fidelity bound'' that follows, we derive a lower bound on $F$ from bounds of the fraction of population $q_{00}$ that remains in the state $|0\rangle_{A}|0\rangle_{S}$ after a beamsplitter acts on $|0\rangle_{A}|0\rangle_{S}$, and the fraction of population $q_{11}$ that remains in the state $|1\rangle_{A}|1\rangle_{S}$ after a beamsplitter acts on $|1\rangle_{A}|1\rangle_{S}$.
Roughly speaking, $1-q_{00}$ provides a measurement of the incoherent displacement errors, and $q_{11}$ provides a measurement of the coherent error.
Before presenting the derivation of the bound, we describe our procedure for inferring $q_{00}$ and $q_{11}$ from our data.

We denote by $p_t(j_A, j_S|k_A, k_S)$ the probability that we measure phonon number $j_A$ in the alternating mode and $j_S$ in the stretch mode after preparing $k_A$ phonons in the alternating mode and $k_S$ phonons in the stretch mode and applying the dynamics for a time $t$.
Let us denote by $t_{\text{BS}}$ the time at which the applied dynamics is the intended 50:50 beamsplitter. In the absence of SPAM error, we have $q_{00} = p_{t_{\text{BS}}}(00|00)$, and similarly for $q_{11}$. However, due to SPAM error, the initial state differs from the intended one, and the measurement outcome may misidentify the final state. We assume the state preparation and measurement procedures are totally dephasing in the number basis. As a result we can describe the preparation, dynamics and measurement processes by stochastic processes. Let \(s(i_{A},i_{S}|k_{A},k_{S})\) be the probability that 
we actually prepare the state \(\ket{i_{A}}_{A}\ket{i_{S}}_{S}\) after the preparation procedure that intends to prepare \(\ket{k_{A}}_{A}\ket{k_{S}}_{S}\).
Let \(m(j_{A},j_{S}|l_{A},l_{S})\) be the probability that the measurement procedure outputs \(j_{A},j_{S}\) when the state before measurement is \(\ket{l_{A}}_{A}\ket{l_{S}}_{S}\). The applied dynamics, with dephasing in the number basis before and after, is described by the probability \(q_t(l_A, l_S|i_A, i_S)\) that the final state is \(\ket{l_{A}}_{A}\ket{l_{S}}_{S}\) when the initial state is \(\ket{i_{A}}_{A}\ket{i_{S}}_{S}\), where \(t\) is the time for which the dynamics are applied. We can model the experimental preparation
and measurement statistics as a function of the duration of the
dynamics as
\begin{align}
    p_t(j_A, j_S|k_A, k_S) = \sum_{i_A, i_S, l_A, l_S}m(j_A, j_S|l_A, l_S)q_t(l_A, l_S|i_A, i_S)s(i_A, i_S|k_A, k_S).
\end{align}
Given this model, we have $q_{00} = q_{t_{\text{BS}}}(00|00)$, and $q_{11} = q_{t_{\text{BS}}}(11|11)$.
At time $t=0$, the dynamics have not acted, so we have
\begin{align}
    q_0(l_A, l_S|i_A, i_S) &= \delta_{i_A, l_A}\delta_{i_S, l_S}\\
    \implies p_0(j_A, j_S|k_A, k_S) &= \sum_{i_A, i_S}m(j_A, j_S|i_A, i_S)s(i_A, i_S|k_A, k_S)
\end{align}
where $\delta_{ab}$ is the Kroenecker delta symbol.
 We approximate
  \(q_{00}\) and \(q_{11}\) under the assumption that the probability
  of two or more preparation, measurement or total-phonon changing
  errors occurring is small enough that it can be neglected.
  \begin{align}
  q_{00}\approx h_{00} &= p_{t_{\text{BS}}}(00|00)/p_0(00|00),\\
  q_{11}\approx  h_{11} &= p_{t_{\text{BS}}}(11|11)/p_0(11|11)
\end{align}
Our bound on the fidelity is correct to first
  order when calculated with \(h_{00}\) and \(h_{11}\) instead of
  \(q_{00}\) and \(q_{11}\).  First order correctness is explained as
  follows: First, the lower bound on fidelity obtained in subsection
  ``Derivation of fidelity bound'' is monotonically increasing in
  \(q_{00}\) and monotonically decreasing in \(q_{11}\). Thus it
  suffices to show that \(h_{00}-q_{00}\) and \(q_{11}-h_{11}\) are
  bounded from above by a second order term. In fact, \(q_{11}-h_{11}\) has
  second order relative error, that is, \((q_{11}-h_{11})/q_{11}\) is
  bounded from above by a second order term. For \(q_{00}\), the
denominator of \(h_{00}\) satisfies
\(p_{0}(00|00)=s(00|00)m(00|00) +O(e^{2})\), where \(O(e^{2})\)
denotes terms quadratic in the errors. In fact,
\(p_{0}(00|00) \geq s(00|00)m(00|00)\), so the second order correction
can only decrease \(h_{00}\) compared to \(q_{00}\).  The numerator is
\(s(00|00)q_{t_{\text{BS}}}(00|00) m(00|00) + O(e^{2})\), so the
ratio yields \(q_{t_{\text{BS}}}(00|00)\) to first order.  For
\(q_{11}\), the denominator of \(h_{11}\) satisfies
\(p_{0}(11|11) = s(11|11)m(11|11) + O(e^{2})\).  The numerator is
\(s(11|11) m(11|11) q_{t_{\text{BS}}}(11|11) + r\), where \(r\) is
positive.  Thus
\(h_{11}\geq q_{t_{\text{BS}}}(11|11) /(1 +
O(e^{2})/(s(11|11)m(11|11)))\). The second order error term can
decrease \(h_{11}\), but only by a factor that is \(1\) to first
order.  Because the quantities \(s(ll|ll)m(ll|ll)\) are \(1\) to first
order, the relative error that can bias our 
bound in the unwanted direction is second order.

It remains to estimate $h_{00}$ and $h_{11}$ from data.
Since measurements are performed only at specific time points \(\{t_{i}\}\)
  and these points did not include \(t_{\text{BS}}\), it is necessary to fit
  the observed measurement outcome probabilities to infer \(t_{\text{BS}}\)
  and their values at \(t_{\text{BS}}\).
At each time point $t_i$, we construct the estimators
$\hat{p}_{t_i}(11|11), \hat{p}_{t_i}(00|00)$ from their corresponding frequencies.
In particular, to estimate the frequency for the 11 outcome, we use only the data where we use mapping sequence 2 in Fig.~\ref{fig:MapSeq}\textbf{b} to read out the alternating mode, and the mapping sequence in Fig.~\ref{fig:MapSeq}\textbf{c} to read out the stretch mode.
We take the number of times that the readout of the alternating mode gave the bright outcome and the readout of the stretch mode gave the outcome that 1 phonon was present, then divide by $N_{rep}$, the number of times that we performed the joint measurement, to obtain the frequency $\hat{p}_{t_i}(11|11)$.
The frequency of the 00 outcome is estimated using the data where we use mapping sequence 1 in Fig.~\ref{fig:MapSeq}\textbf{a} to read out the alternating mode.
It is computed similarly, by counting the number of times that the readout of the alternating mode gave the bright outcome and the readout of the stretch mode gave the outcome that 0 phonons were present, then divide by $N_{rep}$ to obtain the frequency $\hat{p}_{t_i}(00|00)$.

We fit the ratios $\{\hat{p}_{t_i}(11|11)/\hat{p}_{0}(11|11)\}$ to the model
\begin{align}
  g_{11}(t|\tau, A, f, C) = e^{-t/\tau}\left(\frac{1+A \cos(2\pi f t)}{2} + C\right)
  \label{eq:sinusoidalfit}
\end{align}
with $\tau, A, f$ and $C$ as free parameters, using a weighted least squares fit to 21 points, 20 of which were uniformly spaced in time from 29 $\mu s$ to 328 $\mu s$, and one of which was at time $t=0$.
Then we extract $t_{\text{BS}}$ as $t_{\text{BS}} = 1/(2f)$, and our estimator $\hat{h}_{11} = \hat{q}_{11}= g_{11}(t_{\text{BS}}|\tau, A, f, C)$ using the fit values of $\tau, A, f$ and $C$. 
Our fit yielded the values of $\tau = 2.3(6)~\mathrm{ms}$, $A = 0.97(3)$, $f = 0.01016(1)~\mathrm{MHz}$, $C = 0.00(1)$.
We similarly fit the ratios $\{\hat{p}_{t_i}(00|00)/\hat{p}_{0}(00|00)\}$ to the model
\begin{align}
  g_{00}(t|\sigma) = e^{-t/\sigma}
  \label{eq:exponentialfit}
\end{align}
and extract our estimator
$\hat{h}_{00} = \hat{q}_{00} = g_{00}(t_{\text{BS}}|\sigma)$ using
the fit value of $\sigma$.
This model was fit on the same time grid of 21 points.
Our fit yielded the value of $\sigma = 4.7(4)~\mathrm{ms}$.
We then plug these estimators of $q_{00}$
and $q_{11}$ into the bound we derive below for the average fidelity
$F$. 
To obtain a confidence lower bound on $F$, we perform nonparametric bootstrap resampling of the frequencies $\hat{p}_{t_i}(11|11), \hat{p}_{t_i}(00|00)$ and perform the whole fitting procedure 1000 times to obtain a bootstrap distribution of lower
bounds on $F$, then use the bias-corrected percentile
method~\cite{efron1994introduction}.  This procedure gives us a
68\% confidence lower bound of $97.9 \%$ for the average
fidelity of the beamsplitter on the one-phonon subspace.
\newcommand{\hqoo}{0.9896}\newcommand{\hqii}{0.012}
From this procedure, we found a point estimate of $\hat{q}_{00} = \hqoo$ with a 68\% bootstrap confidence interval of $(0.9862, 0.9926)$, and $\hat{q}_{11} = \hqii$, with a 68\% bootstrap confidence interval of $(0.0091, 0.0197)$.

\subsection{Derivation of fidelity bound}
\label{sec:derivation}


To obtain the bound on the fidelity of the beamsplitter operation on the one-phonon subspace, we assume that the noisy beamsplitter is a Gaussian process~\cite{Weedbrook_2012}  that consists of a passive linear optical unitary followed by displacement noise.
We here produce a lower bound on the entanglement fidelity $F_e$~\cite{Nielsen_2002} in terms of $q_{00}$ and $q_{11}$,
which implies a bound on the fidelity \(F\) from the identity $F = (2F_e+1)/3$.

We take the convention that script letters such as $\cO, \cU, \cN$ are
superoperators that act on the space of density matrices.  We model
the noisy beamsplitter $\cO$ as a passive linear optical
transformation $\cU$ followed by random displacement noise $\cN$, so
$\cO = \cN \circ \cU$.  The Gaussian displacement noise is assumed to have has zero
  mean, with the displacements drawn according to a Gaussian
  distribution with $4 \times 4$ covariance matrix
$N$. We represent the displacement vector
  $r=(\Re(x), \Im(x), \Re(y), \Im(y))$ as the concatenation of the real
  and imaginary parts of the two complex phase space displacements  \(x\) and \(y\) of
  the two modes.  The probability distribution of \(r\) is given by
\begin{align}
  \mu(r) &= \frac{1}{4\pi^2\sqrt{\det(N)}}\exp\left( -\frac{1}{2}r^T N^{-1} r \right).
  \label{eq:noisedistn}
\end{align}
The corresponding superoperator acting on density matrix \(\rho\) is
\begin{equation}
  \cN(\rho) = \int dr \mu(r) D_{x,y} \rho D^\dagger_{x,y}.
\end{equation}
This displacement noise model accounts for independent heating noise on the two modes modified by the simultaneously acting dynamics, but also captures more general noise models such as correlated heating on the two modes.
We use $\hat a$ and $\hat b$ for the annihilation operators on the two modes.
Displacement operators satisfy \(D_{-z}=D_{z}^{\dagger}\), \(D_{x,y}\hat a= (\hat a-x)D_{x,y}\),
\(D_{x,y}\hat b= (\hat b-y)D_{x,y}\) and
\(\bra{00}D_{r}\ket{00} = e^{-|r|^{2}/2}\).

We denote by $U$ the linear optical unitary that acts on the mode space.
The general form of \(U\) has phases, but since our measurements are phase insensitive, we can simply adopt a convention for these phases. This corresponds to pre and post multiplying \(U\) by
diagonal unitaries
and modifying \(\cN\) accordingly. With this
  assumption and phase correction, \(U\) can be assumed to be of the
  form
  \begin{align}
    U&=U(\theta)=\begin{pmatrix}
      \cos(\theta+\pi/4) & \sin(\theta+\pi/4)\\
      -\sin(\theta+\pi/4) & \cos(\theta+\pi/4)
    \end{pmatrix},
  \end{align}
where we use the convention that
  \(\theta=0\) is the desired balanced beamsplitter.
\(U\) transforms 
mode operators according to  \(a\mapsto \cos(\theta+\pi/4) a - \sin(\theta+\pi/4) b\) and \(b\mapsto \sin(\theta+\pi/4)a + \cos(\theta+\pi/4)b\).
  Without loss of generality, we can ensure that \(\cos(\theta+\pi/4)\geq 0\) and \(\sin(\theta+\pi/4)\geq 0\),
  which limits \(\theta\)
  to \(\theta\in[-\pi/4,\pi/4]\). We can write \(U(\theta)=\tilde U(\theta) U(0)\),
  where
  \begin{align}
    \tilde U(\theta)&=\begin{pmatrix}
      \cos(\theta) & \sin(\theta)\\
      -\sin(\theta) & \cos(\theta)
    \end{pmatrix}.
  \end{align}
  We refer to \(\tilde U(\theta)\) as the erroneous  passive linear unitary.

Our strategy to compute the bound proceeds in four steps.
{\setlist{nolistsep}
  \begin{itemize}[itemsep=0.5em]
  \item[1.] Bound \(\tr(N)\) in terms of \(q_{00}\).
  \item[2.] Express \(F_{e}\) in terms of \(\theta\), \(q_{00}\) and \(N\).
  \item[3.] Upper bound $\sin^2(\theta)$ in terms of $q_{00}$ and $q_{11}$.
  \item[4.] Simplify by appropriate bounding techniques, eliminate \(\theta\)
    to bound \(F_{e}\) from below in terms of \(q_{00}\), \(q_{11}\) and \(N\),
    and eliminate the dependence on \(N\)  by bounding the surviving terms
    in terms of \(q_{00}\).
  \end{itemize}}
  For step 1, we compute from
  \(q_{00}(r)=|\bra{00}D_{x,y}\ket{00}|^{2}= e^{-r^{T}r}\) that
  \begin{align}
    q_{00} &= \int dr\mu(r) q_{00}(r)\\
           &= \frac{1}{4\pi^{2}\det(N)^{1/2}}
             \int dr e^{-\frac{1}{2} r^{T}(N/(1+2N))^{-1}r}\\
           &=\det(1+2N)^{-1/2}.\label{eq:moment1}
  \end{align}
  The second line in this computation shows that
    \(\mu(r) q_{00}(r) = \det(1+2N)^{-1/2}\nu(r)\), where \(\nu(r)\)
    is a Gaussian probability distribution with mean \(0\) and
    covariance matrix \(M=N/(1+2N)\). Below, we need the following bound on
    the moments of \(\nu(r)\):
    \begin{align}
      \int dr \nu(r) |r|^{2k} \leq (2k-1)!! \tr(M)^{k},
    \end{align}
    where \((2k-1)!!= (2k-1)(2k-3)\ldots\) is the number of perfect
    matchings on \(2k\) elements, and $k\ge 1$. For \(k=1\), the
    inequality is an equality, in particular
    \(\int dr \mu(r)q_{00}(r)|r|^{2} = \det(1+2N)^{-1/2}\tr(M) =
    q_{00}\tr(M)\).  The proof of this fact is provided in the
    subsection ``Bound on moments'' below.

Eq.~(\ref{eq:moment1}) yields
    a bound on \(\tr(N)\) in terms of \(q_{00}\). Since \(N\)
    is positive,
    \begin{align}
      \tr(M) = \tr(N/(1+2N)) \leq \tr(N) \leq (\det(1+2N)-1)/2=(1/q_{00}^2-1)/2.
    \end{align}
To simplify writing the bound, we define \(Q_{00}=(1/q_{00}^2-1)/2\).

For the next step, we require some overlaps
  between number states and displaced number states. Consider a
generic mode \(\hat c\) and start by expressing
\(\bra{1}D_{z}\ket{0}\), \(\bra{1}D_{z}\ket{1}\) and
\(\bra{1}D_{z}\ket{2}\) in terms of \(\bra{0}D_{z}\ket{0}\).
  \begin{align}
    \bra{1}D_{z}\ket{0} &= \bra{0}\hat c D_{z}\ket{0}\\
                        &=\bra{0}D_{z}(\hat c+z)\ket{0}\\
                        &=z\bra{0}D_{z}\ket{0}.\\
    \bra{1}D_{z}\ket{1} &=\bra{0}D_{z}(\hat c+z)\ket{1}\\
                        &=\bra{0}D_{z}\ket{0} +z\bra{0}D_{z}\ket{1}\\
                        &=\bra{0}D_{z}\ket{0}(1-z z^*).\\
    \bra{1}D_{z}\ket{2} &= \frac{1}{\sqrt{2}}\bra{1}D_{z}\hat c^{\dagger}\ket{1}\\
                        &=\frac{1}{\sqrt{2}}\bra{1}(\hat c^{\dagger}-z^*) D_{z}\ket{1}\\
                        &=\frac{1}{\sqrt{2}}(\bra{0}D_{z}\ket{1}- z^*\bra{1}D_{z}\ket{1})\\
                        &=\frac{1}{\sqrt{2}}\bra{0}D_{z}\ket{0}
                          (-z^*-z^*(1-zz^*))\\
                          &=\frac{1}{\sqrt{2}}\bra{0}D_{z}\ket{0}(z^*(\abs{z}^2-2)).
  \end{align}
  \MKs{Here we used the identity \(D_{z}^{\dagger}= D_{-z}\).}
  \MKc{Added this to the list of displacement properties above.}

  Now we perform step 2. The performance of a quantum operation can be characterized by measuring its effect on one half of a state that is maximally entangled with a hypothetical ``reference system,'' which is a copy of the physical system. Specifically, the entanglement fidelity \(F_e\) is the fidelity of the state that is produced by applying the actual operation to one half of a Bell state with respect to the state that is produced by applying the ideal operation to one half of a Bell state. Define \(\ket{d}=\ket{10}\) and \(\ket{u}=\ket{01}\).  The
  standard Bell state is
  \(\ket{\Psi} = (\ket{uu}+\ket{dd})/\sqrt{2}\)  where the second
    system consists of two reference modes.  For a fixed 
  displacement $D_{x,y}$ in the mixture that makes $\cN$, the
  entanglement fidelity is
  $F_e(r) = \lvert O_e(r) \rvert^2 = \lvert \bra{\Psi} (D_{x,y}\tilde
  U(\theta)\otimes I )\ket{\Psi} \rvert^2$, where we abuse notation to also denote the Hilbert space operator corresponding to the erroneous linear optical process by $\tilde{U}(\theta)$.
  The overlap $O_e(r)$ is a sum of four terms, which after multiplying by \(2\) are
  \begin{align}
    \cos(\theta) \bra{d}D_{x,y}\ket{d}\\
    \cos(\theta)\bra{u}D_{x,y}\ket{u}\\
    \sin(\theta)\bra{u}D_{x,y}\ket{d}\\
    -\sin(\theta)\bra{d}D_{x,y}\ket{u}.
  \end{align}
  Adding them up, substituting identities, and accounting for the factor of \(2\) gives
  \begin{align}
    O_{e}(r)&=\bra{00}D_{x,y}\ket{00}\left(
    \cos(\theta)(1-(xx^*+ yy^*)/2)
   + i\sin(\theta) \Im(xy^*)
    \right).
  \end{align}
  The entanglement fidelity is obtained by integrating the product of 
  \begin{align}
    F_{e}(r)= |O_{e}(r)|^{2} &= q_{00}(r)\left(\cos^2(\theta)(1-|r|^{2}/2)^{2} + \sin^2(\theta)\Im(xy^*)^{2}\right)
  \end{align}
  and \(\mu(r)\). 
Note that the second term is positive and high order, so we can drop it since we are computing a lower bound on \(F_{e}\). 
That is, we have the inequalities
  \begin{align}
    F_{e}(r) &\geq q_{00}(r)\cos^2(\theta)(1-|r|^{2}/2)^{2}\\
             &\geq q_{00}(r)\cos^2(\theta)(1-|r|^{2}).\\
    F_{e} &= \int dr \mu(r) F_{e}(r) \\
             &\geq q_{00}\cos^{2}(\theta)(1-\tr(M))\\
             &\geq q_{00}(1-Q_{00})(1-\sin^{2}(\theta)).\label{eq:felb}
  \end{align}
  The last expression is monotonically increasing in \(q_{00}\).

  Now we perform step 3, finding an upper bound on $\sin^2(\theta)$ in terms of $q_{00}$ and $q_{11}$ that can be substituted into Eq.~(\ref{eq:felb}).  The state after \(U\) is
    \(\ket{\psi}=-\sin(2\theta)\ket{11}+\cos(2\theta)(\ket{20}-\ket{02})/\sqrt{2}\).
  We have $q_{11} =
  \bra{11}\cN(\ketbra{\psi})\ket{11}$, which is an integral over
  \(q_{11}(r)=|\bra{11}D_{x,y}\ket{\psi}|^{2}\).
  We can write out the matrix element explicitly as
  \begin{align}
    O_{11}(r) &= \bra{11}D_{x,y}\left(-\sin(2\theta)\ket{11}+\frac{1}{\sqrt{2}}\cos(2\theta)(\ket{20}-\ket{02})\right)\\
           &=\bra{00}D_{x,y}\ket{00}\left(
             -\sin(2\theta)(1-|x|^{2})(1-|y|^{2})
             +\frac{1}{2}\cos(2\theta)\left(
             x^* y |x|^{2}-2 x^*y - x y^* |y|^{2}+2 x y^*
             \right)\right)\\
           &=\bra{00}D_{x,y}\ket{00}\left(
             -\sin(2\theta)(1-|x|^{2})(1-|y|^{2})
             + \frac{1}{2}\cos(2\theta)\left(
            \Re( x^* y)(|x|^{2} - |y|^{2})
            + i \Im(x^* y)(|x|^{2}+|y|^{2} - 4)
             \right)
             \right).
  \end{align}
  For a bound of \(q_{11}(r)\) from below, we can drop
the imaginary part of \(O_{11}(r)\) and, after squaring, the square of the the real
  part of the second summand, which is high order in \(r\). We then obtain the bound
  \begin{align}
    q_{11}(r)  &\geq
                      q_{00}(r)
                      \left(
                      \sin^{2}(2\theta)(1-|x|^{2})^{2}(1-|y|^{2})^{2}
                      + \sin(2\theta)\cos(2\theta)
                           (1-|x|^{2})(1-|y|^{2})\Re( x^* y)(|y|^{2}-|x|^{2})
                      \right)\\
                    &\geq
                      q_{00}(r)
                      \left(
                      \sin^{2}(2\theta)(1-2|x|^{2})(1-2|y|^{2})
                      + \sin(2\theta)\cos(2\theta)
                           (1-|x|^{2})(1-|y|^{2})\Re( x^* y)(|y|^{2}-|x|^{2})
                      \right)\\
                    &\geq
                      q_{00}(r)
                      \left(
                        \sin^{2}(2\theta)(1-2\abs{r}^{2})
                      + \sin(2\theta)\cos(2\theta)
                           (1-|x|^{2})(1-|y|^{2})\Re( x^* y)(|y|^{2}-|x|^{2})
                      \right)\\
                    &\geq
                      q_{00}(r)
                      \left(
                        \sin^{2}(2\theta)(1-2\abs{r}^{2})
                      - \abs{\sin(2\theta)\cos(2\theta)}
                           f(x,y)g(x,y)
                      \right),
  \end{align}
where \(f(x,y)=\abs{(1-|x|^{2})(1-|y|^{2})}\)
  and \(g(x,y)=\abs{\Re(x^* y)(|y|^{2}-|x|^{2})}\). To proceed, we separately bound \(f(x,y)\) and \(g(x,y)\) in terms of polynomials in \(r^{2}\). We fix \(s=r^{2}=|x|^{2}+|y|^{2}\) and write \(u=|x|^{2}/s\in [0,1]\).  Then
  \(f(x,y) = \abs{(1-su)(1-s+su)}\) and \(g(x,y)\leq s^{2}\sqrt{u(1-u)}\abs{2u-1}\).
  We will first consider \(g(x,y)\). The quantity \(\sqrt{u(1-u)}\abs{2u-1}\) is invariant
  under the transformation \(u\mapsto 1/2-(u-1/2)\), which reflects \(u\)
  around the point \(u=1/2\). It therefore suffices to find the maximum
  of \(p(u)=\sqrt{u(1-u)}(1-2u)\) for \(u\in [0,1/2]\). We have \(p(0)=p(1/2)=0\)
  and \(p(u)\geq 0\) on \([0,1/2]\). The derivative of \(\log p(u)\) is
  \begin{align}
    \frac{d}{du}\log p(u) &=
                            \frac{1}{2u} - \frac{1}{2(1-u)} -\frac{2}{1-2u}\\
                          &=\frac{1}{2u(1-u)(1-2u)}
                            \left((1-u)(1-2u)-u(1-2u) - 4u(1-u)\right)\\
                          &=\frac{1}{2u(1-u)(1-2u)}\left(1-8u+8u^{2}\right).
  \end{align}
  This is zero at \(u_{0}=(1-1/\sqrt{2})/2\), and
  \(p(u_{0})=1/4\). Thus \(g(x,y) \leq r^4/4\). At a fixed value of \(s\), the maximum for \(f(x,y)\)  is given by
  \(\max(\max_{u\in[0,1]}(1-su)(1-s+su), - \min_{u\in[0,1]}(1-su)(1-s+su))\).
  We have \((1-su)(1-s + su) = 1 -s + s^{2}u- s^{2}u^{2}\), whose maximum
  is at \(u=1/2\), with value \((1-s/2)^{2}\). The minimum is at the boundary
  and given by \(1-s\). The polynomial \(1-s+s^{2}/2\) is greater than
  both \((1-s/2)^{2}\) and \(s-1\), achieving equality with $s-1$ at $s=2$ and equality with $(1-s/2)^2$ at $s=0$. However, to ensure the monotonicity properties
  for the lower bound on the fidelity, we need a polynomial with positive coefficients,
  so we use \(1+s^{2}/4\) instead.
  Putting together the bounds on \(f(x,y)\)
  and \(g(x,y)\) we get
  \begin{align}
    q_{11}(r)  &\geq  q_{00}(r)
                      \left(
                        \sin^{2}(2\theta)(1-2\abs{r}^{2})
                      - \frac{1}{4}\abs{\sin(2\theta)\cos(2\theta)}
                           (1+r^{4}/4)r^{4}
                      \right).
  \end{align}

  We have \(q_{11}=\int dr\mu(r) q_{11}(r)\). Integrating,
  substituting the bounds on the moments of \(\nu(r)\) and the bound
  \(Q_{00} \geq \tr(M)\) gives
  \begin{align}
    q_{11} &\geq q_{00}\left(
             \sin^{2}(2\theta)(1-2\tr(M))
             -|\sin(2\theta)|\cos(2\theta)
             \left(
             (3+ 105 \tr(M)^{2}/4)\tr(M)^{2}/4
             \right)
             \right)\\
     &\geq q_{00}\left(
             \sin^{2}(2\theta)(1-2Q_{00})
             -|\sin(2\theta)|\cos(2\theta)
             \left(
             (3+ 105 Q_{00}^{2}/4)Q_{00}^{2}/4
             \right)
             \right).
  \end{align}
  Since the quantities \(q_{00}\), \(q_{11}\) and \(Q_{00}\) are determined from
  the experiment, the inequality just obtained can be used to find an upper bound
  on \(\sin^{2}(\theta)\) with \(\theta\in [-\pi/4,\pi/4]\), and this upper bound
  can be substituted into the inequality for \(F_{e}\) obtained earlier.
  For this, note that the expression is symmetric in \(\theta\),
  so it suffices to consider \(\theta\in[0,\pi/4]\). To simplify the expression, we define \(x=\sin^{2}(\theta)\),
  \(a=q_{00}(1-2 Q_{00})\) and \(b=(3+105 Q_{00}^{2}/4)Q_{00}^{2}/4\). Both \(q_{11}\) and \(Q_{00}\) are small
  quantities in our experiment, which means that \(a\) is positive and close to \(1\)
  and \(b\) is positive and close to \(0\).
  Substituting double angle formulas, we need to find the maximum
  \(x\in [0,1/2]\) for which
  \begin{align}
    q_{11}&\geq 4 a x(1-x)- 2b \sqrt{x}\sqrt{1-x }(1-2 x)\\
          &= \sqrt{x(1-x)}(4a \sqrt{x(1-x)} +4b x -2b). \label{eq:q11bnd}
  \end{align}
  Let \(w(x)\) denote the right hand side of this inequality.  We
  have \(w(0)=0\) and \(w(1/2)=a\). The quantity
  \(v(x) = 4a\sqrt{x(1-x)}+4b x-2b\) is monotonically increasing and is negative at \(x=0\).  As a result, \(w(x)\) is strictly
  monotonically increasing in \(x\) when \(v(x) > 0\). If
  \(f(1/2) < q_{11}\), we set \(\sin^{2}(\theta_{\max})=1/2\) in the lower
  bound Eq.~(\ref{eq:felb}) for \(F_{e}\).  If not, then there is a
  unique \(x_{0}\in[0,1/2]\) such that \(w(x_{0})=q_{11}\) and for
  \(x>x_{0}\), we have \(w(x)>q_{11}\). In this case, we set
    \(\sin^{2}(\theta_{\max})=x_{0}\). We can now argue that the bound on \(F_{e}\) is monotonically increasing in \(q_{00}\) and
  decreasing in \(q_{11}\). Because \(Q_{00}\) is monotonically
  decreasing in \(q_{00}\), the quantity \(a\) in Eq.~(\ref{eq:q11bnd})
  is monotonically increasing in \(q_{00}\), and the quantity \(b\) is
  monotonically decreasing.  This implies that
  \(\sin^{2}(\theta_{\max})\) is monotonically decreasing in
  \(q_{00}\). The argument for how \(\sin^{2}(\theta_{\max})\) is
  determined also shows that \(\sin^{2}(\theta_{\max})\) is
  monotonically increasing in \(q_{11}\). It follows by inspection of
  Eq.~(\ref{eq:felb}) that \(F_{e}\) behaves as claimed.

    Plugging in our bootstrap estimates for $\hat{q}_{00}$ and $\hat{q}_{11}$ yields a one-sided 84\% confidence lower bound on entanglement fidelity of $0.968$, and therefore a corresponding 84\% confidence lower bound on average fidelity of $0.979$. We chose this confidence level so that one can directly compare our confidence lower bound to the lower end of a two-sided $68\%$ confidence interval.


    \subsection{Bound on moments}
      \begin{lemma}
    Let \(R\) be a \(\mathbb{R}^{n}\)-valued Gaussian random variable  with
    covariance matrix \(M\) and mean \(0\). Then
    the moments \(\lang (R^{T}R)^{k}\rang\) satisfy
    \begin{align}
      \lang (R^{T}R)^{k}\rang &\leq (2k-1)!!\tr(M)^{k},
    \end{align}
    where \((2k-1)!!=(2k-1)(2k-3)\ldots\)  is the number of perfect matchings
    on a \(2k\)-element set.
  \end{lemma}

  \begin{proof}
    The moments do not change under an orthogonal rotation \(O\) replacing
    \(M\) with \(O^{T}M O\). We can therefore assume that \(M\) is diagonal.
    We prove the inequality by induction on the dimension \(n\). For \(n=1\), 
    \(\lang (R^{T}R)^{k}\rang\) is the \(2k\)-th moment of the Gaussian
    distribution with variance \(M_{11}\) and mean \(0\), which is
    given by \((2k-1)!! M_{11}^{k}\).
    Suppose that the inequality holds for dimension \(n-1 \geq 1\). We prove it for dimension
    \(n\). Let \(S=(R_{1},\ldots, R_{n-1})\) be the projection of \(R\) onto the first
    \(n-1\) coordinates. Let \(N\) be the \(n-1\) by \(n-1\) projection of \(M\) onto
    the first \(n-1\) coordinates.
    Because \(M\) is diagonal, \(S\) is independent of \(R_{n}\).
    Consequently
    \begin{align}
      \lang (R^{T}R)^{k}\rang &= \sum_{j=0}^{k}\binom{k}{j}
                                    \lang (S^{T}S)^{j}R_{n}^{2(k-j)}\rang \nonumber\\
      &\leq \sum_{j=0}^{k}\binom{k}{j} 
                                    (2j-1)!! \tr(N)^{j}(2(k-j)-1)!!M_{nn}^{k-j}.
    \end{align}
    Because \((2j-1)!! (2(k-j)-1)!!\) is the number of perfect matchings
    made up of a perfect matching on the first \(2j\) elements, and another
    on the last \(2(k-j)\) elements, which counts a subset of all perfect matchings,
    \((2j-1)!!(2(k-j-1))!!\leq (2k-1)!!\). Therefore
    \begin{align}
      \lang (R^{T}R)^{k}\rang &\leq (2k-1)!!\sum_{j=0}^{k}\binom{k}{j}
                                    \tr(N)^{j}M_{nn}^{(k-j)}\nonumber\\
      & = (2k-1)!! \tr(M)^{k}.
    \end{align}
  \end{proof}

  From the proof of the lemma, it can be seen that the inequality is saturated
  whenever \(M\) is rank \(1\). It is also saturated for \(k=1\).

\section{Minimization of Mg$^{+}$ participation in the Stretch mode}
\noindent Anharmonicity in the external trap potential and higher-order terms (beyond the term linear in ion displacement) of the Taylor expansion of Coulomb force can impact mode participation of a multi-ion string~\cite{home2011normal}. In our experiments, we consider three types of anharmonic terms including the radial gradients $\partial U / \partial x$ and $\partial U / \partial y$, the twist curvature term $\partial^{2} U / (\partial x \partial z)$, and the axial cubic term  $\partial^{3} U/ \partial z^{3}$. Non-zero values for these terms lead to non-zero participation of the Mg$^{+}$ ion in the Stretch mode, which will then be coupled to recoil from Mg$^{+}$ scattering events. We cool the Stretch mode to its ground state and apply ``shim" voltages (shims) calculated from trap simulations to mimimize the relevant anharmonic terms. Experimentally we calibrate the applied shim strengths by minimizing the heating of the Stretch mode from Mg$^+$ photon scattering. Radial gradients are controlled by a differential voltage shim on the pair of control electrodes closest to the ions and an additional voltage shim applied to a bias electrode on a third-layer wafer outside the two main wafers~\cite{blakestad2010transport}.
The twist term is controlled by the two pairs of electrodes next to the two electrodes closest to the ions. By applying a differential voltage shim of $v$ on the pair of electrodes on one side and $-v$ on the other, the twist of the ion string in the $x-z$ plane relative to the trap axis can be minimized. 
The cubic term $z^3$ is controlled with the same electrodes used for creating the Alternating-Stretch coupling. 
The optimal shim values are determined as follows: All three axial modes are sideband cooled close to their ground states, then the voltage shim controlling one of the anharmonicity terms is set to a certain strength. Mg$^{+}$ resonant light is pulsed for 1\,ms to cause scattering on Mg$^{+}$, then the applied anharmonicity shim is set back to zero and probability of a state change of Be$^{+}$ when driving a MSS pulse is determined. The shim with minimal MSS spin-flip probability is retained as optimal and applied for subsequent experiments.  Since the shims are not perfectly decoupled from each other, we iterate the minimization of all three anharmonic terms for several rounds to find the overall best shims.
However, we still observe small residual heating of the Stretch mode due to scattering on Mg$^{+}$. This could be because the anharmonicity is still not totally eliminated. Another potential source is radiation pressure on the Mg$^{+}$ ion, which shifts its position relative to the Be$^+$ ions slightly during photon scattering and breaks the mirror symmetry of the ion string.
\\
\\
\section{Calibration of the Cirac-Zoller mapping for repeated motional state measurements}
\noindent In the repeated motional state measurements, a Cirac-Zoller (CZ) sequence maps information about the Alternating mode state onto the Mg$^{+}$ internal state. The basic principle is that a motion-subtracting-sideband 2$\pi$ pulse does not change motional states $|0\rangle$ and $|1\rangle$ but leads to a motional-state-dependent phase shift on the Mg$^{+}$ internal state. The Mg$^{+}$ is prepared in $\left|3,1\right\rangle_{M}$ with optical pumping, followed by a microwave $\pi$ pulse sequence. Then, a microwave carrier $\pi$/2 pulse of $\left|3,1\right\rangle_{M} \leftrightarrow \left|2,0\right\rangle_{M}$ generates a superposition state $1/\sqrt{2}\left(\left|2,0\right\rangle_{M}+\left|3,1\right\rangle_{M}\right)$. Next, the population in $\left|3,1\right\rangle_{M}$ is transferred to $\left|2,2\right\rangle_{M}$ by a microwave $\pi$ pulse followed by a MSS 2$\pi$ pulse of $\left|2,2\right\rangle_{M}\left|1\right\rangle_{A} \leftrightarrow \left|3,3\right\rangle_{M}\left|0\right\rangle_{A}$. For $\left|n=0\right\rangle_{A}$, the MSS pulse does not change the Alternating mode or the Mg$^{+}$ internal state, and the system remains in $1/\sqrt{2}(\left|2,0\right\rangle_{M}+\left|2,2\right\rangle_{M})\left|0\right\rangle_{A}$. For $\left|n=1\right\rangle_{A}$, the MSS pulse drives $\left|2,2\right\rangle_{M}\left|1\right\rangle_{A}$ to $\left|3,3\right\rangle_{M}\left|0\right\rangle_{A}$ and back to $-\left|2,2\right\rangle_{M}\left|1\right\rangle_{A}$, flipping the sign of this component. The state is changed to $1/\sqrt{2}(\left|2,0\right\rangle_{M}-\left|2,2\right\rangle_{M})\left|1\right\rangle_{A}$ in this case. After the MSS pulse, the population in $\left|2,2\right\rangle_{M}$ is transferred back to $\left|3,1\right\rangle_{M}$ with a microwave $\pi$ pulse. Subsequently, a second microwave $\pi$/2 pulse of $\left|3,1\right\rangle_{M} \leftrightarrow \left|2,0\right\rangle_{M}$ with a relative phase $\phi_{2}$ with respect to the first $\pi$/2 pulse is applied. The populations in $\left|3,1\right\rangle_{M}$ and $\left|2,0\right\rangle_{M}$ are shelved to $\left|3,3\right\rangle_{M}$ (the bright state) and $\left|2,-2\right\rangle_{M}$ (the dark state) respectively by a microwave sequence before Mg$^{+}$ fluorescence detection.
\\
\\
\noindent The probability of Mg$^{+}$ being measured in bright ($b$) or dark ($d$) is given by $P(b)= (1-\cos(\phi_{2}))/2$ and $P(d)= (1+\cos(\phi_{2}))/2$. By setting $\phi_{2} = 0$, $\left|0\right\rangle_{A}$ is mapped to $d$ while $\left|1\right\rangle_{A}$ is mapped to $b$. The mapping $\left|0\right\rangle_{A} \rightarrow b$ and $\left|1\right\rangle_{A} \rightarrow d$ is realized by setting $\phi_{2} = \pi$.  The particular transitions chosen in the CZ mapping implementation yield the shortest duration of the sequence in our apparatus, which reduces errors due to heating and dephasing. The phases $\phi_{2}$ are calibrated periodically to account for experimental drifts.  We calibrate $\phi_{2}$ with the Alternating mode prepared in $\left|0\right\rangle_{A}$ and $\left|1\right\rangle_{A}$, which yields two out-of-phase sinusoidal signals as a function of $\phi_2$, as shown in Fig.~\ref{fig:fig12}\textbf{a}; we use the $\phi_{2}$ yielding maximal or minimal fluorescence for realizing the respective mappings. When the Alternating mode contains population in states with $\left|n>1\right\rangle_{A}$, the MSS pulse no longer accomplishes a 2$\pi$ rotation and changes the Alternating mode state. Therefore, this particular measurement does not preserve the motional state if it contains any population outside $\{\left|0\right\rangle_{A}, \left|1\right\rangle_{A}\}$.
\\
\\
\section{State verification after repeated motional state measurements}
\noindent The final state after $N$ motional state measurements is examined with MAS and MSS $\pi$ pulses. In Fig.~\ref{fig:fig12}\textbf{b}, with $N=1$ and M1, we detect $\{o_{1}\}=\{d\}$ heralding $\left|0\right\rangle$ with a probability of 0.960(3), close to $p_0$ but with a small difference indicating a readout error of about $\epsilon_{0}$\,$\approx$\,0.02. 
The MAS (pink bar) and MSS (violet bar) results conditioned on $\{d\}$ are close to their ideal values (hatched bars) for $\left|0\right\rangle$, suggesting that this conditioned final state is preserved during the measurement and close to $\left|0\right\rangle$. However, the sideband results conditioned on $\{o_{1}\}=\{b\}$ heralding $\left|1\right\rangle$ significantly deviate from their ideal values for $\left|1\right\rangle$. This is because the probability of getting a false outcome for $\left|0\right\rangle$, $p_{0}\epsilon_{0}\approx$\,0.02, is comparable to $p_{1}$, thus causing a noticeable effect on the heralded results for $\left|1\right\rangle$. Erroneous declarations can be reduced by repeating the measurement. When $N=2$, the sideband results conditioned on $\{o_{1},o_{2}\}=\{b, b\}$ are significantly closer to the ideal expectation for $\left|1\right\rangle$ even though Mg$^{+}$ scatters twice as many photons. When $N=3$, the final states conditioned on $\{o_{1},o_{2},o_{3}\}=\{b, b, b\}$ match the expectation values even more closely. The post-selected results for $\left|0\right\rangle$ also show improved accuracy relative to the MAS/MSS analysis when increasing the number of measurements. We observed similar performance when implementing M2 but with the roles of $\left|0\right\rangle$ and $\left|1\right\rangle$ flipped. 
\\
\\
\section{Characterization of heating associated with repeated motional state measurements}
\noindent We perform a series of tests to characterize the heating from each element in the motional state measurement sequence. All three axial modes are sideband cooled to near their ground states and the Mg$^{+}$ ion is prepared in $\left| \downarrow \right\rangle_{M}$. Then, modified measurement sequences are implemented with different combinations of CZ mapping, swap pulses, Mg$^{+}$ detection, and Mg$^{+}$ sideband cooling of the in-phase (INPH) and Alternating modes. Elements that are not applied are sometimes replaced with a delay of equal duration to account for anomalous heating. This is followed by either applying a Alternating MAS or MSS $\pi$ pulse and determining the respective spin-flip probabilities $P_{\rm MSS}$ and $P_{\rm MAS}$. The ratio of spin-flip probabilities $r=P_{\rm MSS}/P_{\rm MAS}$ is then used to estimate the average motional occupation $\bar{n}=r/(1-r)$ of the Alternating mode. This estimate is accurate if the motional state is well-described by a thermal distribution or if the probability for number states with $n>1$ can be neglected. More details and results of these tests are shown in Table~\ref{tab:QND_nbar_test}.
\\
\\
We measure the mean occupation of the Alternating mode immediately after sideband cooling to be $\bar{n}_{\rm SBC}$\,=\,0.023(1) as a reference value. In the ``no swaps" test, the two swaps are replaced with two delays of equal duration, which yields $\bar{n}$\,=\,0.040(3), higher than $\bar{n}_{\rm SBC}$ by $\Delta\bar{n}$\,=\,0.017(3) due to the heating of the Alternating mode acting over the delay replacing the second swap. This test also approximately sets a lower bound for $\bar{n}$ when at least one measurement is performed and no outcome is used for post-selection. 
Next, we perform the swap test where only two swap pulses are applied after sideband cooling. We find a rise in $\bar{n}$ by $\Delta\bar{n}$\,=\,0.021(4) from two swaps.
In the CZ heating test, a delay of equal duration as a CZ mapping sequence is inserted before swaps and further increases $\bar{n}$ by $\Delta\bar{n} \approx$\, 0.005.
To estimate the $\Delta\bar{n}$ due to heating of the Stretch mode during Mg$^{+}$ detection and SBC of the other two modes, we apply a delay equal to the duration of Mg$^{+}$ detection followed by Mg$^{+}$ sideband cooling and find an increase of $\Delta\bar{n}$\,=\,0.010(5).
In the recoil heating test, we apply Mg$^{+}$ detection to investigate the additional heating of the Stretch mode from recoil of photons scattered on Mg$^{+}$. We estimate $\Delta\bar{n}\approx$\,\, 0.012(5) added to $\bar{n}$ from scattered photons alone. 
An ion typically scatters on the order of a few tens of photons during sideband cooling while on the order of 10$^{3}$ photons are scattered during fluorescence detection. Assuming recoil heating is proportional to scattered photon number, scattering during sideband cooling only causes a negligible gain in $\bar{n}$ on the order of 10$^{-4}$.
Lastly, we implement a test (labelled as ``No exchange" in Fig.~\ref{fig:fig3}\textbf{d}) of the main text) that replaces $N$=1,\,2,\,3 measurement blocks with an equal delay and obtain $\bar{n}$\,=\,0.25(3),\,0.51(6),\,0.8(1), respectively, which are significantly higher than $\bar{n}_{\rm  SBC}$, because the motional state resides longer in the Alternating mode at a higher heating rate, compared to $N$ measurement blocks where the mode is swapped to the Stretch mode for substantial parts of the experiments.

\newpage
\begin{table}[h]
    \centering
    \begin{tabular}{c|*3c|*3c}
          \hline
          Exp. & \multicolumn{3}{c|}{M1 ($N$\,=\,1)} & \multicolumn{3}{c}{M2 ($N$\,=\,1)} \\
          \hline
          $\{o_{1}\}$ & $\{d\}$ & $\{b\}$ & Overall & $\{d\}$ & $\{b\}$ & Overall\\
          \hline
          MAS & 0.931(4) & 0.76(3) & 0.924(3) & 0.84(2) & 0.915(4) & 0.910(4)\\
          \hline
          MSS & 0.030(2) & 0.49(3) & 0.048(3) & 0.33(2) & 0.044(3) & 0.064(3)\\
          \hline
          Prob. & 0.960(3) & 0.040(3) & 1 & 0.066(3) & 0.934(3) & 1\\
          \hline
    \end{tabular}
    \caption{\textbf{Sideband transition probabilities conditioned on different motional state measurement outcomes for $N$=1}. The table shows the probability of a spin flip after a MAS or MSS $\pi$ pulse, conditioned on $\{o_{1}\}$=$\{d\}$ and $\{b\}$ for mapping M1 and M2 and the probability of outcomes $d$ and $b$. The overall sideband spin-flip probabilities with no conditioning on measurement outcomes are also shown.}
    \label{tab:QND_SB1}
\end{table}

\begin{table}[h]
    \centering
    \begin{tabular}{c | *{5}{c} | *{5}{c}}
          \hline
          Exp. & \multicolumn{5}{c|}{M1 ($N$\,=\,2)} & \multicolumn{5}{c}{M2 ($N$\,=\,2)} \\
          \hline
          $\{o_{1}, o_{2}\}$ & $\{d,d\}$ & $\{d,b\}$  & $\{b,d\}$ & $\{b,b\}$ & Overall & $\{d,d\}$ & $\{d,b\}$  & $\{b,d\}$ & $\{b,b\}$ & Overall\\
          \hline
          MAS & 0.924(4) & 0.78(2) & 0.88(2) & 0.57(4) & 0.908(4) & 0.56(4) & 0.91(2) & 0.73(2) & 0.929(4) & 0.903(4)\\
          \hline
          MSS & 0.027(2) & 0.46(3)  & 0.08(2) & 0.77(4) & 0.066(3) & 0.73(4) & 0.05(1) & 0.53(2) & 0.044(3) & 0.102(4)\\
          \hline
          Prob. & 0.900(4) & 0.049(3) & 0.029(2) & 0.022(2) & 1 & 0.025(2) & 0.042(3) & 0.083(4) & 0.850(5) & 1\\
          \hline
    \end{tabular}
    \caption{\textbf{Sideband transition probabilities conditioned on different motional state measurement outcomes for $N$\,=\,2}. Similar to table \ref{tab:QND_SB1}, listing results for all combinations of two measurement outcomes.}
    \label{tab:QND_SB2}
\end{table} 

\begin{table}[h]
    \centering
    \begin{tabular}{c|*{11}{c}}
          \hline
          Exp. & \multicolumn{11}{c}{M1 ($N$\,=\,3)} \\
          \hline
          $\{o_{1}, o_{2}, o_{3}\}$ & $\{d,d,d\}$ & $\{d,d,b\}$ & $\{d,b,d\}$ & $\{d,b,b\}$ & $\{b,d,d\}$ & $\{b,d,b\}$  & $\{b,b,d\}$ & $\{b,b,b\}$ & \{Majority $d$\} & \{Majority $b$\} & Overall\\
          \hline
          MAS & 0.935(2) & 0.74(2) & 0.89(2) & 0.64(3) & 0.93(2) & 0.60(8) & 0.75(6) & 0.59(4) & 0.926(2) & 0.63(2) & 0.912(3)\\
          \hline
          MSS & 0.025(2) & 0.51(2) & 0.13(2) & 0.78(2) & 0.03(1) & 0.78(7) & 0.22(6) &  0.82(3) & 0.047(2) & 0.74(2) & 0.08(2)\\
          \hline
          Prob. & 0.876(6) & 0.036(2) & 0.020(1) & 0.024(1) & 0.021(1) & 0.0028(5) & 0.0046(6) & 0.016(1) & 0.952(1) & 0.048(1) & 1\\
          \hline
          \hline
          Exp. & \multicolumn{11}{c}{M2 ($N$\,=\,3)} \\
          \hline
          $\{o_{1}, o_{2}, o_{3}\}$ & $\{d,d,d\}$ & $\{d,d,b\}$ & $\{d,b,d\}$ & $\{d,b,b\}$ & $\{b,d,d\}$ & $\{b,d,b\}$  & $\{b,b,d\}$ & $\{b,b,b\}$ & \{Majority $d$\} & \{Majority $b$\} & Overall\\
          \hline
          MAS & 0.48(9) & 1.0 & 0.8(1) & 0.92(3) & 0.66(5) & 0.88(4) & 0.72(4) & 0.927(7) & 0.64(4) & 0.908(7) & 0.891(7)\\
          \hline
          MSS & 0.80(8) & 0.07(7) & 0.4(2) & 0.03(2) & 0.76(5) & 0.04(2) & 0.46(4) & 0.045(5) & 0.68(4) & 0.077(6) & 0.118(7)\\
          \hline
          Prob. & 0.015(3) & 0.004(1) & 0.006(2) & 0.034(4) & 0.043(5) & 0.042(5) & 0.075(6) & 0.783(9) & 0.067(4) & 0.933(4) & 1\\
          \hline
    \end{tabular}
    \caption{\textbf{Sideband transition probabilities conditioned on different motional state measurement outcomes for $N$\,=\,3.} Similar to the previous two tables. In addition, majority $d$ or $b$ is conditioned on at least two out of three outcomes being $d$ or $b$.}
    \label{tab:QND_SB3}
\end{table}

\begin{table}[h]
    \centering
    \begin{tabular}{c|*4c|*3c}
          \hline
          & CZ mapping & Swaps & Mg detection & Mg SBC & $\bar{n}$ &  $\Delta\bar{n}$\\
         \hline
          SBC & No & No & No & No & 0.023(1) & 0 \\
          \hline
         No-swaps test & Yes & Replaced with delay & Yes & Yes & 0.040(3) & 0.017(3)\\
          \hline
          Swap test & No & Yes & No & No & 0.044(3) & 0.021(4)\\
          \hline
          CZ heating test & Replaced with delay & Yes & No & No & 0.049(4) &  0.026(5)\\
          \hline
          Stretch heating test & Replaced with delay & Yes & Replaced with delay & Yes & 0.059(4) & 0.036(5)\\
          \hline
          Recoil heating test & Replaced with delay & Yes & Yes & Yes & 0.071(5) & 0.048(5)\\
           \hline
         No exchange ($N=1$) & Replaced with delay & Replaced with delay & Replaced with delay & Replaced with delay & 0.25(3) & 0.23(3)\\
         \hline
         No exchange ($N=2$) & Replaced with delay & Replaced with delay & Replaced with delay & Replaced with delay & 0.51(6) & 0.49(6)\\
          \hline
         No exchange ($N=3$) & Replaced with delay & Replaced with delay & Replaced with delay & Replaced with delay & 0.8(1) & 0.8(1)\\
          \hline
    \end{tabular}
    \caption{\textbf{Experiments to characterize heating associated with certain parts of the motional state measurement}. A series of tests to delineate heating of different elements of the measurements, including CZ mapping, swaps, Mg detection, and Mg SBC. Each element is either applied (``Yes"), omitted (``No"), or ``Replaced with delay" of the same duration as the element in these tests. The Alternating mode $\bar{n}$ of all tests and the relative increase $\Delta \bar{n}=\bar{n} - \bar{n}_{\rm  SBC}$ are listed.}
    \label{tab:QND_nbar_test}
\end{table}

\begin{table}[h]
    \centering
    \begin{tabular}{c | *{2}{c} *{3}{c}}
          \hline
          Exp. &  \multicolumn{2}{c|}{M1 ($N$\,=\,1)} & \multicolumn{3}{c}{M1 ($N$\,=\,2)}\\
          \hline
          Condition & $\{o_{1}\}$=$\{d\}$ & \multicolumn{1}{c|}{No post-selection}  & $\{o_{2}\}$=$\{d\}$ & $\{o_{1},o_{2}\}$=$\{d,d\}$ & No post-selection \\
          \hline
          $\bar{n}$& 0.034(3) & \multicolumn{1}{c|}{0.055(3)} & 0.032(3) & 0.030(3) & 0.078(4)\\
          \hline
          \hline
          Exp. &  \multicolumn{5}{c}{M1 ($N$\,=\,3)} \\
          \hline
          Condition & $\{o_{3}\}$=$\{d\}$ & $\{o_{2},o_{3}\}$=$\{d,d\}$  & $\{o_{1},o_{2},o_{3}\}$=$\{d,d,d\}$ & \{Majority $d$\} & No post-selection\\
          \hline
          $\bar{n}$ &  0.032(2) & 0.028(2)  & 0.027(2) & 0.053(2) & 0.095(3)\\
          \hline
    \end{tabular}
    \caption{\textbf{Mean occupations of the Alternating mode conditioned on different motional state measurement outcomes with mapping M1 ($\left|0\right\rangle \rightarrow d$, $\left|1\right\rangle \rightarrow b$)}.}
    \label{tab:QND_nbar_M1}
\end{table}

\begin{table}[h]
    \centering
    \begin{tabular}{c | *{2}{c} *{3}{c}}
          \hline
          Exp. &  \multicolumn{2}{c|}{M2 ($N$\,=\,1)} & \multicolumn{3}{c}{M2 ($N$\,=\,2)}\\
          \hline
          Condition & $\{o_{1}\}$=$\{b\}$ & \multicolumn{1}{c|}{No post-selection}  & $\{o_{2}\}$=$\{b\}$ & $\{o_{1},o_{2}\}$=$\{b,b\}$ & No post-selection \\
          \hline
          $\bar{n}$& 0.050(3) & \multicolumn{1}{c|}{0.075(4)} & 0.049(3) & 0.049(3) & 0.127(6)\\
          \hline
          \hline
          Exp. &  \multicolumn{5}{c}{M2 ($N$\,=\,3)} \\
          \hline
          Condition & $\{o_{3}\}$=$\{b\}$ & $\{o_{2},o_{3}\}$=$\{b,b\}$  & $\{o_{1},o_{2},o_{3}\}$=$\{b,b,b\}$ & \{Majority $b$\} & No post-selection\\
          \hline
          $\bar{n}$ & 0.050(6) & 0.050(6)  & 0.050(6) & 0.093(8) & 0.15(9)\\
          \hline
    \end{tabular}
    \caption{\textbf{Mean occupations of the Alternating mode  conditioned on different motional state measurement outcomes with mapping M2 ($\left|0\right\rangle \rightarrow b$ and $\left|1\right\rangle \rightarrow d$)}.}
    \label{tab:QND_nbar_M2}
\end{table} 

\begin{figure}
    \centerline{\includegraphics[width=1\textwidth]{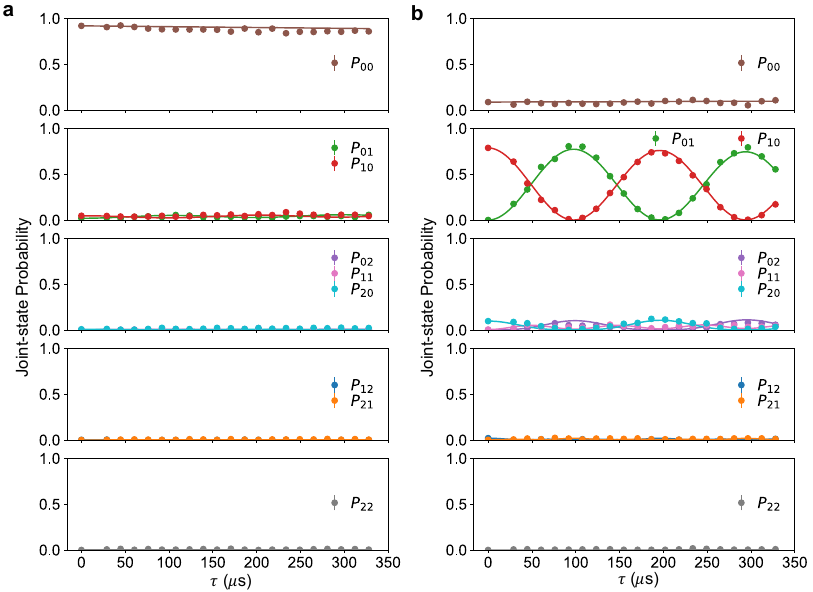}}
    \caption{\textbf{Joint-motional population dynamics: Coupling time scans I.} The two normal modes are prepared in (\textbf{a}) $\left|0\right\rangle_{A}\left|0\right\rangle_{S}$ or (\textbf{b}) $\left|1\right\rangle_{A}\left|0\right\rangle_{S}$. The plots show population in the nine joint number states of the Alternating mode and the Stretch mode as a function of coupling time $\tau$ in five separate panels with (from top to bottom) 0 to 4 total quanta of motion in the two modes. Imperfect state preparation and measurement cause the population of the target initial state to deviate from one at $\tau$\,=\,0 while the other state populations may start with a non-zero value. More detailed descriptions of the experiments can be found in the main text.
     } 
    \label{fig:fig7}
\end{figure}
\begin{figure}
    \centerline{\includegraphics[width=1\textwidth]{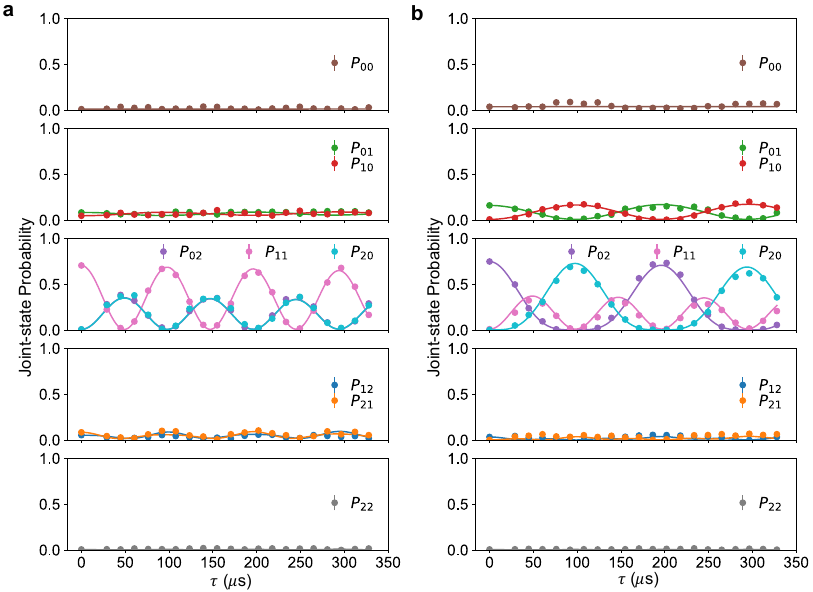}}
    \caption{\textbf{Joint-motional population dynamics: Coupling time scans II.} The Alternating and Stretch modes are prepared in (\textbf{a}) $\left|1\right\rangle_{A}\left|1\right\rangle_{S}$ , (\textbf{b}) $\left|0\right\rangle_{A}\left|2\right\rangle_{S}$.
     } 
    \label{fig:fig8}
\end{figure}

\begin{figure}[h]
    \centerline{\includegraphics[width=1\textwidth]{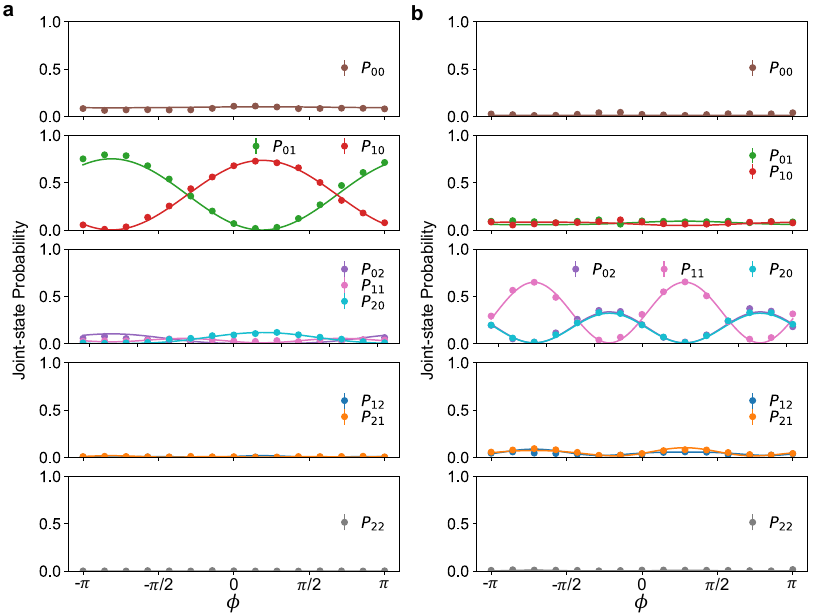}}
    \caption{\textbf{Joint-motional population dynamics: Coupling phase scans.} Phonon interference with the initial states (\textbf{a}) $\left|1\right\rangle_{A} \left|0\right\rangle_{S}$  and (\textbf{b}) $\left|1\right\rangle_{A} \left|1\right\rangle_{S}$.} 
    \label{fig:fig9}
\end{figure}

\begin{figure}[h]
    \centerline{\includegraphics[width=1\textwidth]{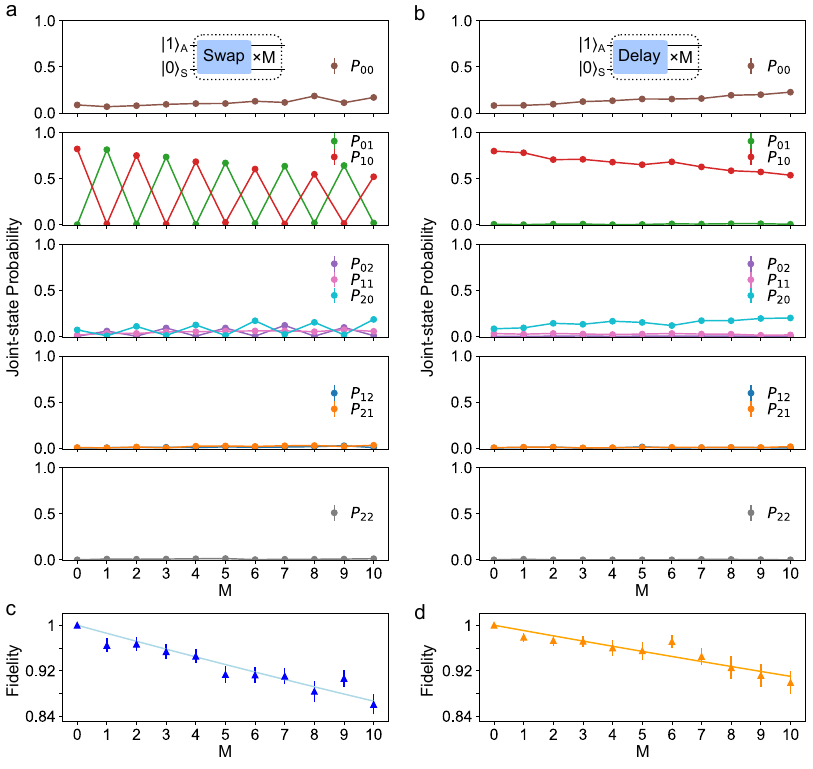}}
    \caption{\textbf{Joint-motional population dynamics during repeated swap operations.} 
    \textbf{a}, With the two modes prepared in  $\left|1\right\rangle_{A}\left|0\right\rangle_{S}$, we apply M swap pulses and measure the populations of the two-mode joint states. The second panel shows that the injected single phonon is swapped between the two modes. The sum of populations in the second panel decreases as the number M of swap operations increases, while the populations of the states shown in the first and third panel grow because of heating. \textbf{b}, Data from a reference experiment where the coupling pulses are replaced with delays of equal duration to illustrate the effect of heating. In this case, no population is exchanged and only a slow population leakage from the initial state (red dots in the second panel) to primarily $\left|0\right\rangle_{A}\left|0\right\rangle_{S}$ (dots in the first panel) and $\left|2\right\rangle_{A}\left|0\right\rangle_{S}$ (blue dots in the third panel) is observed. This is due to a much larger heating rate in the Alternating mode compared to the Stretch mode.  
    Solid lines in \textbf{a} and \textbf{b} serve as guides to the eye. \textbf{c}, With the population of all nine joint states, we estimate the fidelity of the final density matrix $\sigma$ compared to the target density matrix $\rho$ where ideal swaps are applied to an initial density matrix (data at M\,=\,0 in \textbf{a}). We treat the density matrices $\sigma$ and $\rho$ as a fully decohered mixture of nine joint number states (only diagonal terms are non-zero) and estimate the fidelity with $F=(Tr\sqrt{\sqrt{\rho} \sigma \sqrt{\rho}})^{2}$. We fit the fidelities (blue triangles) to $F(\rm M) = (1 - \epsilon)^{\rm M}$ to extract an error $\epsilon$ per swap operation to be 1.4(1)\%. \textbf{d}, The same analysis is performed for the reference experiment data (orange triangles) where the fidelity to the initial density matrix (data at M\,=\,0 in \textbf{b}) is shown and fitted, yielding an error per swap time of 0.9(1)\%. The error of the swap operation is dominated by heating, which can be suppressed by increasing the coupling strength to reduce the swap time, or by lowering the heating rate, for example by operating in a similar trap at cryogenic temperatures.
    }
    \label{fig:fig10}
\end{figure}

\begin{figure*}
    \centerline{\includegraphics[width=1\textwidth]{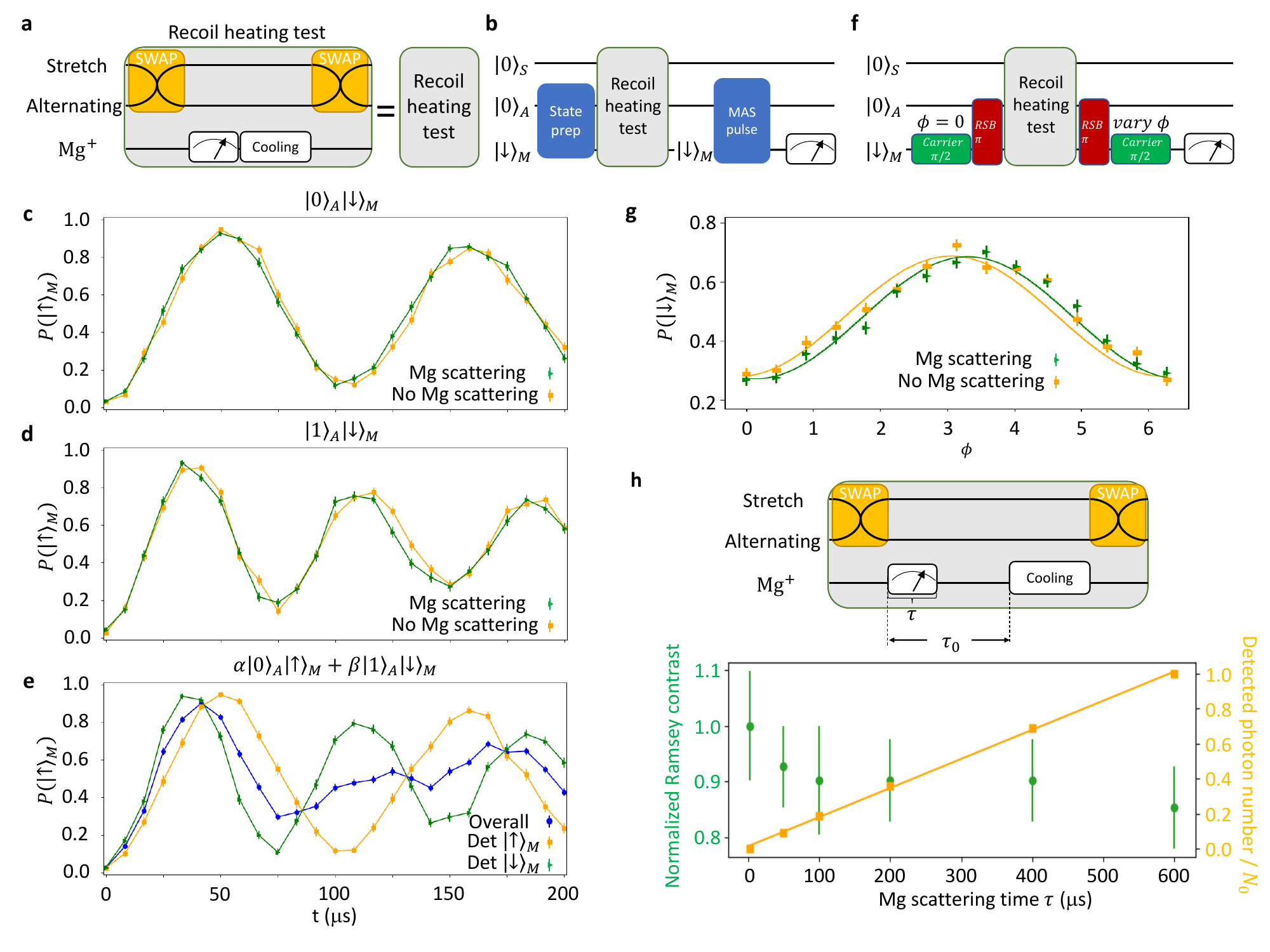}}
    \caption{\textbf{Effects of recoil heating on the protected mode.} \textbf{a}, Circuit diagram for Mg$^{+}$ recoil heating test. The test sequence consists of two swap pulses that exchange the states between the Stretch (``protected") and the Alternating mode, Mg$^{+}$ detection, and a sideband cooling sequence for the INPH mode. Mg$^{+}$ detection can be replaced with a delay of the same duration as required. \textbf{b}, Circuit diagram for testing survival of motional population from Mg$^{+}$ scattering. We prepare the Alternating mode and the Mg$^{+}$ ion in $\left|0\right\rangle_{A}\left|\downarrow\right\rangle_{M}$ (results shown in \textbf{c}), $\left|1\right\rangle_{A}\left|\downarrow\right\rangle_{M}$ (\textbf{d}), and $\alpha \left|0\right\rangle_{A}\left|\uparrow\right\rangle_{M} + \beta \left|1\right\rangle_{A}\left|\downarrow\right\rangle_{M}$ where $|\alpha|^{2} \approx |\beta|^{2} \approx 0.5$ (\textbf{e}), then apply the recoil test sequence with or without Mg$^{+}$ scattering, and finally apply a Alternating MAS pulse with a varying duration followed by Mg$^{+}$ fluorescence detection. In \textbf{c} and \textbf{d}, without (orange squares) and with (green triangles) scattering thousands of photons from the Mg$^+$ ion, we obtained nearly identical MAS excitation traces,  implying that the motional state is not perturbed substantially by the Mg$^+$ ion recoil while stored in the Stretch mode. 
    \textbf{e}, The blue dots are the MAS oscillation results averaged over all experimental trials while disregarding the detection outcome from Mg$^{+}$ in the middle of the experiment, with a rapid decay of contrast due to the mixture of motional states that remains after detection. However, if the MAS results are sorted based on the middle detection outcomes, we obtain two traces for when $\left|\uparrow\right\rangle_M$ and $\left|\downarrow\right\rangle_M$ are detected that project the motional state onto $\left|0\right\rangle_A$ (orange squares) and $\left|1\right\rangle_A$ (green triangles) respectively. The sorted trace for $\left|0\right\rangle_{A}$ ($\left|1\right\rangle_{A}$) is nearly identical to the trace in \textbf{c} (\textbf{d}) where $\left|0\right\rangle_{A}$ ($\left|1\right\rangle_{A}$) is directly prepared without a measurement on Mg$^+$.
    The decay of contrast in these oscillation curves is mainly caused by fluctuations in the Rabi frequency (Debye-Waller effect) due to substantial heating of the in-phase mode.
    \textbf{f}, Circuit diagram for testing survival of motional coherence between $\left|0\right\rangle_{A}$ and $\left|1\right\rangle_{A}$ during Mg$^{+}$ scattering. A carrier $\pi$/2 pulse of $\left| \downarrow\right\rangle_{M}\leftrightarrow\left| \uparrow\right\rangle_{M}$ with a subsequent red-sideband (RSB) $\pi$ pulse of $\left| \downarrow\right\rangle_{M}\left|1\right\rangle_{A}\leftrightarrow\left| \uparrow\right\rangle_{M}\left|0\right\rangle_{A}$ prepares the Alternating mode in a superposition state $1/\sqrt{2}(\left|0\right\rangle_{A}+\left|1\right\rangle_{A})$ and Mg$^{+}$ in $\left| \downarrow\right\rangle_{M}$. After the recoil heating test sequence, another RSB $\pi$ pulse and carrier $\pi$/2 pulse with varying phase $\phi$ relative to the first $\pi$/2 pulse close the motional Ramsey interferometer. 
    \textbf{g}. The motional Ramsey fringes with (green triangles) or without (orange squares) Mg$^{+}$ scattering have similar contrast which indicates the motional coherence is preserved despite scattering many photons while detecting the internal state of Mg$^{+}$. The imperfect Ramsey contrast of these two curves is likely caused by imperfect preparation, motional dephasing and readout errors. 
    \textbf{h}. With a fixed duration $\tau_0$\,=\,800\,$\mu$s between the first swap and recooling of the INPH mode, we vary Mg$^{+}$ scattering time $\tau$. We observe the detected photon number (orange squares, normalized to $N_{0}$\,=\,51.6) linearly increases over $\tau$ while the motional Ramsey contrast (green dots, normalized to the contrast of 0.41(4) for $\tau$\,=\,0\,$\mu$s) remains unchanged within the experimental uncertainty as more photons are scattered.}
    \label{fig:fig11}
\end{figure*}

\begin{figure*}
    \centerline{\includegraphics[width=1.0\textwidth]{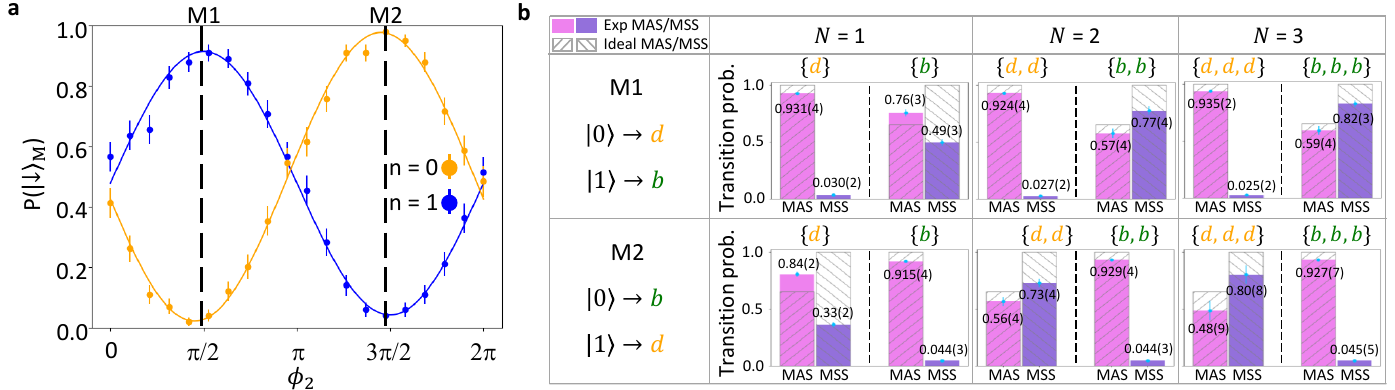}}
    \caption{\textbf{Cirac-Zoller mapping phase calibration and sideband transition probabilities conditioned on different motional state measurement outcomes.} \textbf{a}, With the Alternating mode prepared in $\left|0\right\rangle_{A}$ (orange) or $\left|1\right\rangle_{A}$ (blue), we obtain two sinusoidal curves that oscillate out of phase as a function of the phase $\phi_{2}$ of the second $\pi$/2 pulse relative to the first $\pi$/2 pulse in the CZ sequence. The two vertical dashed lines indicate the phases for realizing mappings M1 and M2. \textbf{b},  MAS and MSS transition probabilities conditioned on the outcomes shown above each panel; measurements are repeated $N$ = 1, 2, 3 times under both M1 and M2. The grey hatched bars denote the ideal probabilities for $\left|0\right\rangle$ and $\left|1\right\rangle$ which are compared with experimental results (colored bars). For M1 (M2), when $|0\rangle$ is heralded by detecting $d$ ($b$) $N$=1, 2, 3 times, the sideband results match well with the ideal values. When $|1\rangle$ is heralded once ($N$=1), the sideband results deviate significantly from the ideal values for $|1\rangle$ because of false declaration events for $|0\rangle$ (see text in Supplementary Material). After performing the measurement more times ($N$=2 and 3), the sideband results when $|1\rangle$ is heralded are substantially closer to the ideal expectation values because the detection error is reduced.
    }
    \label{fig:fig12}
\end{figure*}

\end{document}